\documentclass[lettersize,journal]{IEEEtran}
 \usepackage{amsmath,amssymb}
 \usepackage{subfigure}
 \usepackage{graphicx,graphics,color,psfrag}
 \usepackage{cite,balance}
 \usepackage{algorithm}
 \usepackage{accents}
 \usepackage{amsthm}
 \usepackage{bm}
 \usepackage{url}
 \usepackage{algorithmic}
 \usepackage[english]{babel}
 \usepackage{multirow}
 \usepackage{enumerate}
 \usepackage{cases}
 \usepackage{stfloats}
 \usepackage{dsfont}
 \usepackage{color,soul}
 \usepackage{amsfonts}
  \usepackage{tcolorbox}
\usepackage[utf8]{inputenc}
 \usepackage{cite,graphicx,amsmath,amssymb}
 \usepackage{subfigure}
 \usepackage{fancyhdr}
 \usepackage{hhline}
 \usepackage{graphicx,graphics}
 \usepackage{array,color}
 \usepackage{amsmath}
 \usepackage{amsthm}

\newtheorem{proposition}{Proposition}
\newtheorem{remark}{Remark}
\newtheorem{theorem}{Theorem}
\newtheorem{lemma}{Lemma}

\newtheorem{assumption}{Assumption}

\begin{document}
\include{header}

\title{Joint Sensing, Communication, and Computation for Vertical Federated Edge Learning in Edge Perception Networks}

\author{Xiaowen Cao, Dingzhu Wen, Suzhi Bi, Yuanhao Cui, Guangxu Zhu, Han Hu, and Yonina C. Eldar
\thanks{Xiaowen Cao and Suzhi Bi are with the College of Electronic and Information Engineering, Shenzhen University, Shenzhen  518172, China (email: \{caoxwen,bsz\}@szu.edu.cn). X. Cao is also with Guangdong Provincial Key Laboratory of Future Networks of Intelligence, Shenzhen  518172, China.}

\thanks{Dingzhu Wen is with the School of Information Science and Technology, ShanghaiTech University, Shanghai 201210, China (e-mail: wendzh@shanghaitech.edu.cn). 
}

\thanks{Yuanhao Cui is with the School of Information and Communication Engineering, Beijing University of Posts and Telecommunications, Beijing 100876, China (e-mail: cuiyuanhao@bupt.edu.cn).}
\thanks{Guangxu Zhu is with Shenzhen Research Institute of Big Data, The Chinese University of Hong Kong-Shenzhen, Guangdong, 518172, China (email: gxzhu@sribd.cn).}
\thanks{Han Hu is with the School of Information and Electronics, Beijing Institute of Technology, Beijing 100081, China (e-mail: hhu@bit.edu.cn).}
\thanks{Yonina C. Eldar is with Faculty of Math and CS, Weizmann Institute of Science, Rehovot, Israel (email:yonina.eldar@weizmann.ac.il).}

}
\maketitle

\begin{abstract}
Combining wireless sensing and edge intelligence, edge perception networks enable intelligent data collection and processing at the network edge. However, traditional sample partition based horizontal federated edge learning (HFEEL) struggles to effectively fuse complementary multi-view information from distributed devices. To address this limitation, we propose a vertical federated edge learning (VFEEL) framework tailored for feature-partitioned sensing data.
In this paper, we consider an integrated sensing, communication, and computation (ISCC)-enabled edge perception network, where multiple edge devices utilize wireless signals to sense environmental information for updating their local models, and the edge server aggregates feature embeddings via over-the-air computation (AirComp) for global model training.
First, we analyze the convergence behavior of the ISCC-enabled VFEEL in terms of the loss function degradation in the presence of wireless sensing noise and aggregation distortions during AirComp. Then, to accelerate convergence, we aim to optimize the batch size, sensing power, and transmission power control at edge devices as well as the denoising factors at the edge server under limited network constraints on overall energy consumption and per-round latency. Due to the tight coupling of variables, the problem is non-convex. To address this problem, we design an alternating optimization–based algorithm to efficiently obtain a high-quality solution.
Numerical results are conducted based on a human motion recognition task to verify that the proposed ISCC-enabled VFEEL algorithm achieves higher accuracy compared with other benchmarking schemes including ISCC-enabled HFEEL approach.

\end{abstract}
\vspace{-15pt}

\section{Introduction}

Next-generation networks towards for intelligent applications such as industrial Internet of Things, digital twins, and smart cities, demand high-precision sensing and ultra-low-latency processing \cite{KLetaif19_6G}. To achieve this, wireless sensing can efficiently extract dynamic environmental information \cite{Liu20COMSTWireSen}, but cloud-based data processing may incur high latency and privacy risks.  Edge intelligence mitigates this by deploying local computation capacities at base stations (BSs) and devices. However, this remains limited by passive sensing, which cannot flexibly expand sensing coverage. These challenges drive the \textit{edge perception} paradigm, which integrates wireless sensing and intelligence at the network edge \cite{Cui2024aa,Liu2025EP}. 

\begin{figure}[t]
  \centering
  \includegraphics[width=0.4\textwidth]{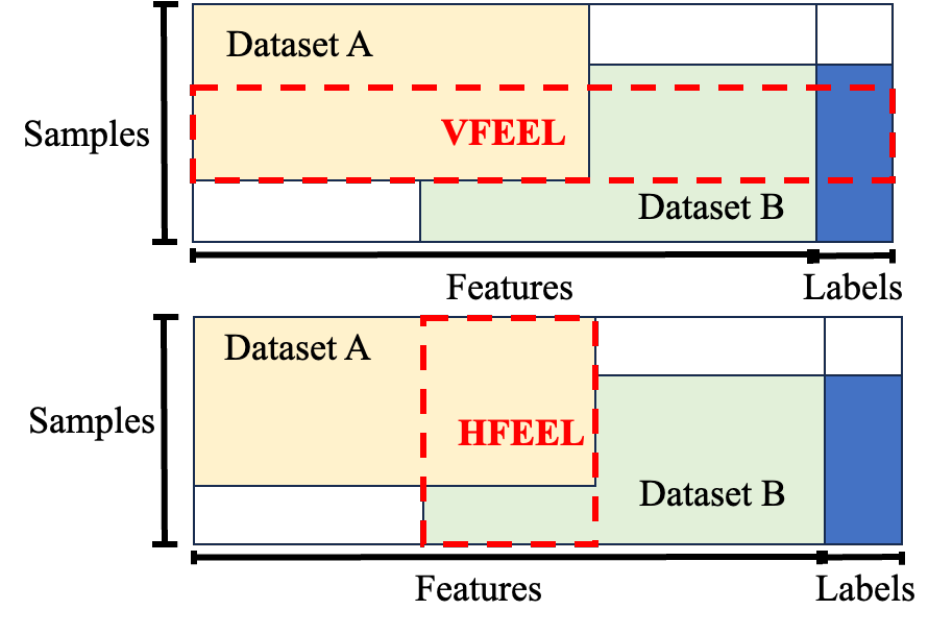}
  \caption{VFEEL versus HFEEL \cite{Yang19FL}.}
  \label{Fig:vfeel}
  \vspace{10pt}
\end{figure}

To enable efficient edge perception, massive sensing data are leveraged to help  artificial intelligence (AI) models understand and adapt to diverse environments. On the one hand, incorporating wireless sensing into existing communication systems has enabled a promising technique called integrated sensing and communication (ISAC) that improves the spectrum utilization efficiency and sensing coverages \cite{FanLiu-JSAC,zhang2025ISAC}. Devices distributed at different locations can provide richer accurate environmental information from various perspectives \cite{Yang2024MVS}. However, how to effectively fuse multi-view sensing data for sequential intelligent processing remains an open challenge due to data heterogeneity and spatial correlation.
On the other hand, federated edge learning (FEEL) has gained significant attention due to its advantages in data privacy and security \cite{Google17FL}. As shown as in Fig. \ref{Fig:vfeel}, FEEL is typically categorized into horizontal FEEL (HFEEL) where data is partitioned by samples across devices with identical features, and vertical FEEL (VFEEL) where data is partitioned by features across devices \cite{Yang19FL,Liu24TKDE}. Although HFEEL efficiently aggregates knowledge across data samples, it struggles to fully exploit the diverse feature representations inherent in multi-view sensing data \cite{PLiu2022_ISACVerFEL}.
This motivates our study on efficient edge perception networks based on VFEEL.

\subsection{Related Work}

The training procedure in FEEL often incurs high communication overhead and latency due to frequent updates between devices and edge servers.  To address this bottleneck, over-the-air FEEL (AirFEEL) has emerged by leveraging over-the-air computation (AirComp) in FEEL, which enables simultaneous model aggregation from multiple devices over a shared spectrum, thus reducing communication and latency \cite{Cao-ComM}. 
To improve learning performance, extensive research was explored ranging from device selection \cite{KYang2020TWC}, power control optimization \cite{Cao2021-FedAvg,Cao2021AirFEEL}, interference mitigation \cite{Wang24TWC_AirComp}, to differential privacy \cite{DLiu2020Ar}. 
While these studies have extensively addressed the performance bottlenecks of HFEEL, they cannot be directly applied to VFEEL case due to the inherent incompleteness of local models in edge devices \cite{Yang19FL}. Recent works in \cite{ZengICC23,ShiTWC24} had made preliminary attempts by considering AirComp enabled two-layer VFEEL, where power control based on channel inversion is used to align intermediate prediction results (i.e. embeddings) across devices. Yet, they overlooked the coupling of data collection and processing, and have not explored how computation errors induced by limited device resources as well as channel and sensing noises affect learning performance.

In ISAC, researchers have sought to quantify ISAC performance limits from the perspectives of capacity–distortion Pareto boundary \cite{Caire18ISIT}, Cramér–Rao bound (CRB) for target estimation \cite{Liu21TSP_CRB}, and CRB-rate tradeoff for bi-static case \cite{Song25CRB}. Unlike most existing work considering a single-link scenario that only captures limited information, distributed wireless sensing nodes can observe the same target from different views which provides diverse features to describe it. This observation has been captured and applied in recognition \cite{ChenGRSL18} and communication \cite{Tong23ComMag}. 
To deal with the heterogeneous data generated from multi-view sensing, \cite{HuangTVT24} employed multi-node collaborative sensing to offload high-precision sensing data to edge servers, thus improving sensing accuracy at the cost of privacy and resource demands. Alternatively, \cite{LiuCL22} proposed a VFEEL framework, where multi-view sensing was used for feature alignment, thereby increasing the precision of recognition tasks. Although these works demonstrated the potential of multi-view sensing for efficient data acquisition, how to efficiently feed them into the VFEEL framework with theoretical analysis for improving learning accuracy is still challenging.


Building upon the advantages of AirFEEL and ISAC, the integrated sensing, communication, and computation (ISCC) framework has been proposed to unify both paradigms which enables jointly design efficient data sensing and FEEL architectures \cite{FengISCC22}, thereby enhancing distributed edge learning \cite{Liu-JSTSP2023} and inference \cite{Wen23TWC_Inference,Li23TWC_CoInference}. 
Specifically, edge devices are able to wirelessly sense the objects for collaboratively training a learning model under the coordination of an edge server for recognition tasks, while AirComp is adopted to facilitate fast gradient aggregation among devices \cite{WenTWC25}.
Existing work mainly focused on device scheduling \cite{DuIoT24}, resource optimization \cite{QiTCom22,WenTWC25}, and sensing strategies \cite{Cai2025aa}. In particular, \cite{QiTCom22} optimized beamforming to balance system performance among three while aggregation error was analyzed in \cite{DuIoT24}, both of which failed to treat learning performance as the core optimization objective. Our recent work advanced this by characterizing aggregation errors from channel and sensing noises, and then jointly optimizing resource allocation and sensing strategies to accelerate convergence \cite{WenTWC25}. Moreover, a task-oriented sensing strategy was proposed in \cite{Cai2025aa} for automatically adapting to training progress to reduce generalization error. 
Current works mainly characterize the performance limits of HFEEL, which are not directly transferable to VFEEL due to the inherent incompleteness of local models at edge devices \cite{Yang19FL,Liu24TKDE}. 

\subsection{Contribution}

In this paper, we proposed an ISCC-based VFEEL framework for edge perception to fully explore the multi-view sensing data from distributed edge devices which are coordinated by an edge server to collaboratively train a recognition model. Specifically, in each round, edge devices use wireless signals to sense targeted objects for updating their local model. Then, they upload the intermediate prediction results (instead of raw data or model parameter/gradient) via AirComp to the edge sever for global model updating. 
Although the convergence of ISCC-enabled AirFEEL has been mathematically characterized in \cite{WenTWC25}, which reveals the impact of sensing and channel noises on HFEEL performance, it does not align well with the feature-partition property of distributed sensing data, as it fails to exploit the feature diversity from multi-view sensing observations. In other words, how to evaluate the impact of aggregation and sensing noises on convergence performance on VFEEL in the presence of incomplete local models remains unexplored, leaving a gap in theoretical guidance for designing ISCC scheme in resource-constrained networks. In addition, the tight coupling of sensing, communication, and computation processes also compounds this gap, as they compete for the same limited resources. This thus motivates our work, and the detailed contributions are listed below. 

\begin{itemize}
\item {\bf ISCC-based VFEEL Framework for Edge Perception}: We first establish a practical ISCC-based VFEEL framework in edge perception network that elaborates on the processes of sensing for data acquisition, on-device computation for local embeddings execution, and AirComp for embeddings aggregation. Particularly, edge devices use wireless signal to sense targeted objects and pre-process the raw data (such as, via data cleaning, data augmentation, filter, etc.\cite{Li23TWC_CoInference}) to feed into the learning model for computing embeddings. We model the sensing noise with respect to (w.r.t.) each sample and characterize the aggregation error induced by AirComp. 

\item {\bf Convergence Analysis}: We first capture the impact of aggregation errors (i.e., the bias and mean squared error (MSE) of the embedding aggregation) on the convergence performance of the ISCC-based VFEEL algorithm based on the first-order Taylor approximation of the training loss function. It is proved that the convergence is accelerated with a larger total batch size at each round for accessing more data samples into training. Unlike the insight that involving more devices to increase learning performance in horizontal AirFEEL \cite{WenTWC25,Cao2021-FedAvg}, it reveals that more edge devices participating may slow down the convergence since the induced sensing and aggregation errors would degrade the local model updating.

\item {\bf Resource Allocation}: 
Building on the convergence analysis, we aim to jointly optimize the batch size, the sensing power, and the transmission power control at edge devices as well as the denoising factors at the edge server to achieve fast convergence under limited network constraints on overall energy consumption and per-round latency. Due to the tight coupling of variables, the problem is non-convex and hard to solve optimally. To address this, we develop an alternating optimization based algorithm to efficiently obtain a high-quality solution.

\item {\bf Performance Evaluation}: 
Finally, we conduct numerical simulations based on a human motion recognition task \cite{LiuCL22} to evaluate the performance of ISCC-based VFEEL system. It is validated that the proposed scheme can achieve higher testing accuracy than other baseline approaches under the same delay and energy budgets as it jointly optimizes batch size and network resources to fully exploit the interplay among sensing, communication, and computation.

\end{itemize}

\section{System Model}

\begin{figure}[t]
  \centering
  \includegraphics[width=0.43\textwidth]{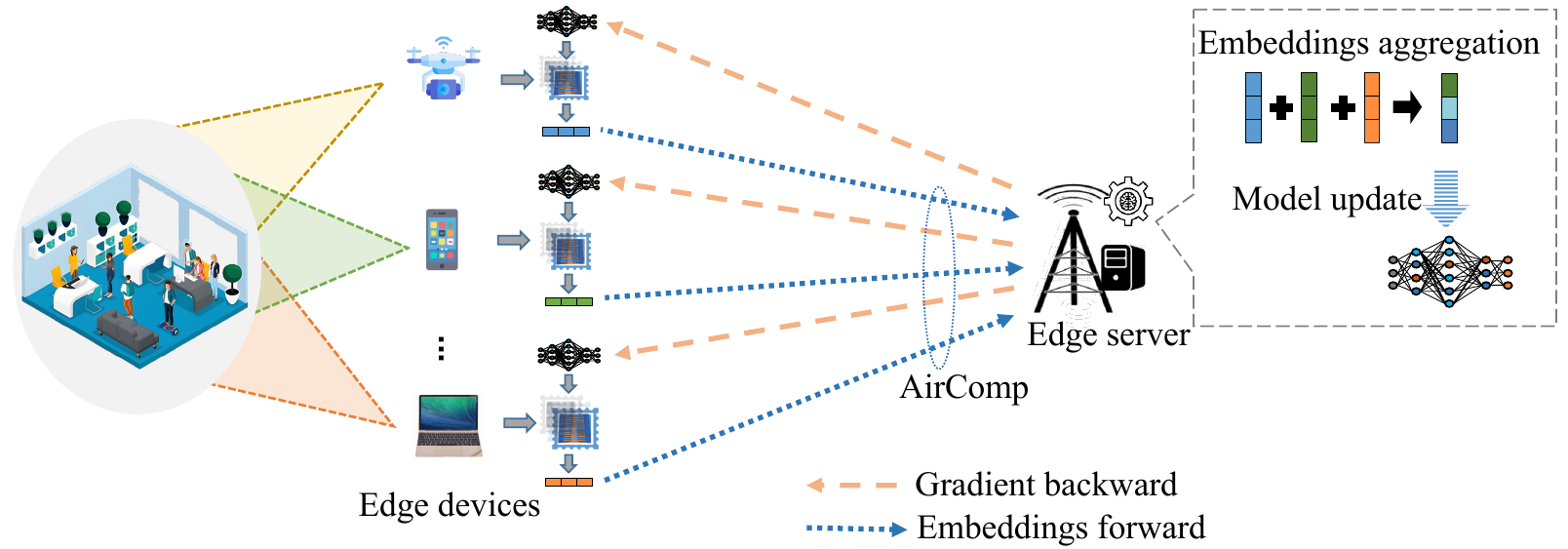}
  \caption{Illustration of ISCC-enabled VFEEL system.}
  \label{Fig:System}
  \vspace{10pt}
\end{figure}

We consider an ISCC-enabled VFEEL system, as illustrated in Fig. \ref{Fig:System}, where $K$ edge devices are coordinated by an edge server to collaboratively train a shared machine learning model. 
Each edge device is equipped with a single-antenna ISAC transceiver. Thus it endows a mode shifting between wireless sensing and communication in a shared radio-frequency circuit by adopting a time-division approach. 
In the sensing mode, edge devices collect sensing data through wireless signals by processing received echo signals for local model training. Simultaneously, all local predictions (embeddings) are uploaded via AirComp-based aggregation for global updates.  The following sections will introduce the V-FEEL algorithm, the data sensing model, and the AirComp-based embeddings aggregation scheme.



\subsection{Vertical Federated Edge Learning Algorithm}


The focus of V-FEEL is to collaboratively train a global machine learning model under the coordination of an edge server.
Suppose that edge device $k$ has its learning model with parameters vector denoted as ${\bm \theta}_{k}\in\mathbb{R}^{V_k}$ with $V_k$ denoting the number of elements, and a local embedding function denoted by $\psi_k (\cdot), \forall k\in\mathcal{K}\triangleq \{1,...,K\}$.
Let $\mathcal{P}_k$ denote the local data distribution at edge device $k$ and ${\bm \xi}_{k,i}\sim\mathcal{P}_k$ represent a random variable following the distribution $\mathcal{P}_k$ whose realization corresponds to a data sample at edge device $k\in \mathcal{K}$. The sample datasets across different edge devices contain disjoint subsets of features (i.e., feature-partitioned data). 
Let ${\bm \xi}_{i}=\left[ {\bm \xi}_{1,i},\cdots,{\bm \xi}_{K,i}\right]$ be the $i$-th complete sample and $y_i$ denote the label of the $i$-th training sample. It is assumed that all labels are available at the edge server.
For ease of illustration, we define $\mathcal{P}$ as the overall data distribution across all edge devices, while $\mathcal{P}_k$ refers to the local view from the $k$-th edge device.

The edge server trains a central model with parameters denoted by ${\bm \theta}_{0}\in\mathbb{R}^{V_0}$, and has a fusion model ${\psi}_0 (\cdot)$ used for collecting all embeddings from all edge devices. Typical fusion scheme includes sum, element-wise averaging, and concatenation \cite{PLiu2022_ISACVerFEL,ceballos2020splitn}.
In this work, the fusion function at the edge server focuses on taking a sum of the embeddings for a sample as input and conducting a predicted label.  
Defining  $f_i(\cdot)$ as a sample-wise loss function, the objective in V-FEEL is to minimize a loss function as follows 
\begin{align}\label{Sys-LearningPro}
	\min_{{\bm \Theta}} ~ F\left({\bm \Theta}\right)=\mathbb{E}_{{\bm \xi}_{i}\sim\mathcal{P}} f_i\left({\bm \theta}_0; {\psi}_i ({\bm \theta}_1,\cdots,{\bm \theta}_K;{\bm\xi}_i)\right),
\end{align}
where ${\bm \Theta}=\left[ {\bm \theta}_{0},\cdots,{\bm \theta}_{K}\right]\in\mathbb{R}^{V}$ is global model with dimensions $V$ and $ {\psi}_i ({\bm \theta}_1,\cdots,{\bm \theta}_K;{\bm\xi}_i)=\sum\limits_{k\in\mathcal{K}}  {\psi}_k ({\bm \theta}_k;{\bm \xi}_{k,i})$ represents the summation of embeddings of all edge devices.



\begin{figure}[t]
  \centering
  \includegraphics[width=0.45\textwidth]{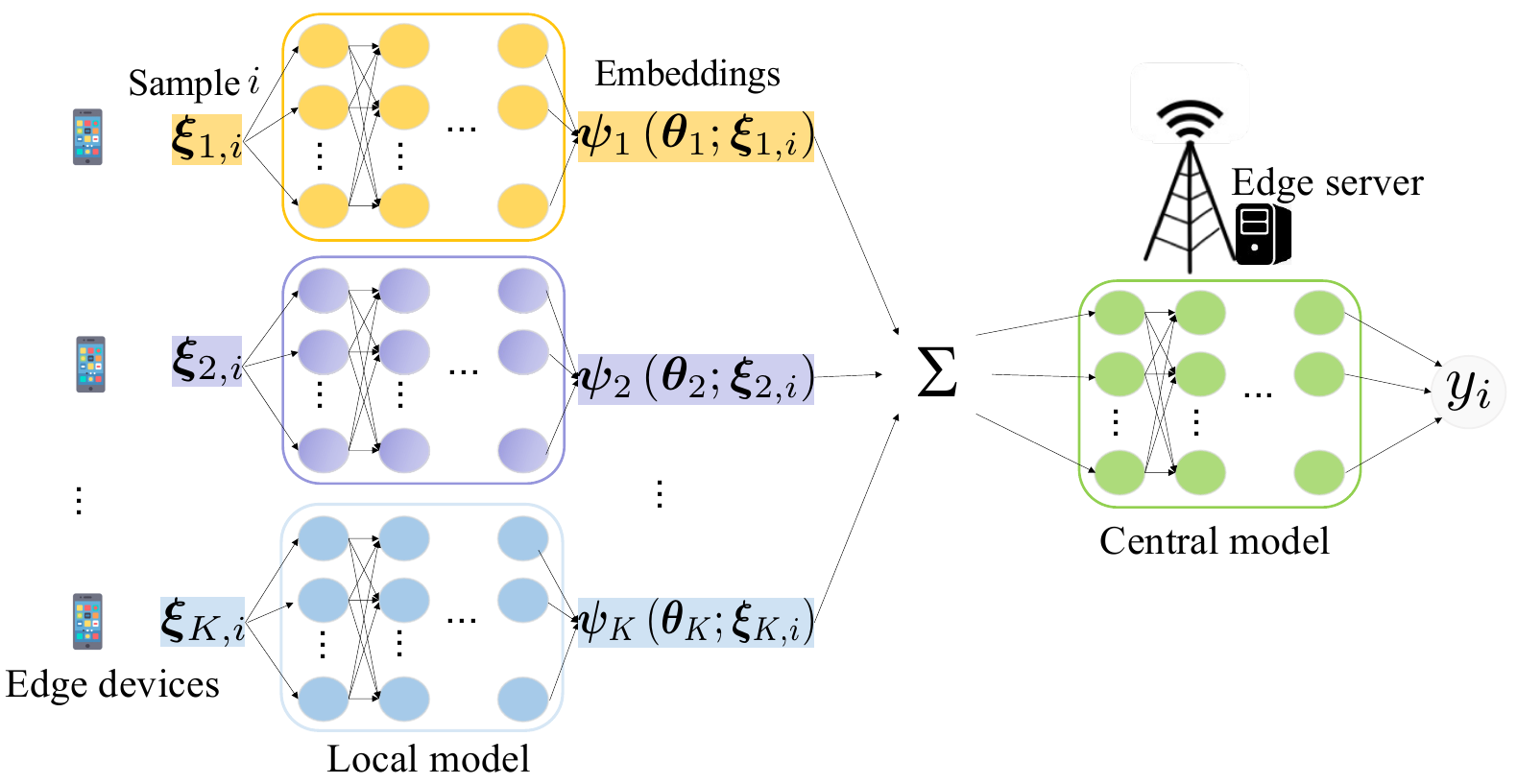}
  \caption{Example local view of a global model in V-FEEL at each round.}
  \label{Fig:Model}
  \vspace{10pt}
\end{figure}

To solve problem \eqref{Sys-LearningPro} while preserving the data privacy for each device, we adopt the distributed \textit{stochastic gradient decent} (SGD) algorithm in V-FEEL, which is implemented iteratively in a distributed manner as follows.
The whole training process includes multiple communication rounds, each of which involves both the forward propagation for loss function evaluation and the backward propagation for gradient updating. 
Besides, each edge device completes a single local SGD on its own local model parameter ${\bm \theta}_k$.
Next, we take any arbitrary round $t\in\mathcal{T}\triangleq\{1,\cdots,T\}$ as an example to illustrate the training process, as depicted in Fig. \ref{Fig:Model}. 
\begin{itemize}
	\item{\bf Training Samples Collection}: Each edge devices collect a batch of noisy sensing data samples denoted by $\mathcal{B}_{k}^{(t)}\triangleq\{\tilde{\bm \xi}_{k,i}^{(t)}\}_{i=1}^{b^{(t)}}$ for local training through wireless sensing in the $t$-th round\footnote{All edge devices are assumed to simultaneously sense the same target from different views, ensuring a synchronized process where the resulting distributed sensing data are naturally aligned. }, where $b^{(t)}$ represents the batch size of sensing data at edge devices, and can be adaptively adjusted over different rounds. 

\item {\bf Local Computation Phase}: Edge device $k$ would input each noisy data sample $\tilde{\bm \xi}_{k,i}^{(t)}$ to the local model ${\bm \theta}_{k}^{(t)}$ for obtaining an embedding ${\psi}_{k} \left( {\bm \theta}_{k}^{(t)};\tilde{\bm \xi}_{k,i}^{(t)}\right)$, $\forall i\in\mathcal{B}_{k}^{(t)}$. 

	\item  { \bf Embedding Forward Phase}: Each edge device forwards its embedding ${\psi}_{k} \left( {\bm \theta}_{k}^{(t)};\tilde{\bm \xi}_{k,i}^{(t)} \right)$ in the meantime to the edge server through AirComp for fast aggregation. Let $\tilde{\bm \psi}_{i}^{(t)} := \tilde{\psi}_i \left({\bm \theta}_1^{(t)},\cdots,{\bm \theta}_K^{(t)};\tilde{\bm \xi}_{i}\right)$ denote the estimate embedding received at the edge server. 
	 Thus, the edge server could get a predicted output for data sample $i$ as $\tilde{ y}_{i}^{(t)}={ \psi}_0^{(t)}\left({\bm\theta}_0^{(t)};\tilde{\psi}_i \left({\bm \theta}_1^{(t)},\cdots,{\bm \theta}_K^{(t)};\tilde{\bm \xi}_{i}\right) \right)$.
	 The sample-wise loss function is 
	 \begin{align}\label{Sys_LossCal}
	 f_i\left({\bm\theta}_0^{(t)};\tilde{\bm \psi}_{i}^{(t)} \right)&=\varepsilon_0\left(\tilde{ y}_i^{(t)},{ y}_i^{(t)} \right),
	 \end{align}
	 where $\varepsilon_0(\cdot)$ denotes the error function between the ground-truth and predicted values, such as cross entropy loss or mean squared error. 

	\item {\bf Gradient Backward Phase}: With the obtained sample-wise loss function, 
the gradient of the central model at the edge server is 
\begin{align}
\hat{\bm g}\left({\bm \theta}_{0}^{(t)}\right)&=	\frac{1}{b^{(t)}}\sum_{i=1}^{b^{(t)}} \nabla_{{\bm \theta}_{0}^{(t)}} f_i\left({\bm\theta}_0^{(t)};\tilde{\bm \psi}_{i}^{(t)} \right)\notag\\
&=\!\frac{1}{b^{(t)}}\!\!\sum_{i=1}^{b^{(t)}}\! \nabla_{\!{\bm \theta}_{0}^{(t)}}\tilde{ \psi}_{0,i}^{(t)}\nabla_{\tilde{ \psi}_{0,i}^{(t)}} f_i\!\left(\!{\bm\theta}_0^{(t)};\!\tilde{\bm \psi}_{i}^{(t)} \!\right),\label{Sys_ChainRule_Server}
\end{align}
where $ \nabla  f\left( \cdot\right)$ denotes the gradient of $f\left( \cdot\right)$
and for national convenience, we have $\tilde{ \psi}_{0,i}^{(t)}:={ \psi}_0^{(t)}\!\left(\!{\bm\theta}_0^{(t)};\tilde{\bm \psi}_{i}^{(t)} \right)$.
Besides, to obtain the gradient of each local model ${\bm \theta}_{k}^{(t)}$ denoted by $ \hat{\bm g}\left({\bm \theta}_{k}^{(t)}\right)$ at edge device $k\in\mathcal{K}$, it needs to be calculated via the chain rule as in \eqref{Sys_ChainRule}.
\begin{figure*}
\begin{align}
&\hat{\bm g}\left({\bm \theta}_{k}^{(t)}\right) =\frac{1}{b^{(t)}}\sum_{i=1}^{b^{(t)}} \nabla_{{\bm \theta}_{k}^{(t)}} f_i\left({\bm\theta}_0^{(t)};\tilde{\bm \psi}_{i}^{(t)} \right)=\frac{1}{b^{(t)}}\!\sum_{i=1}^{b^{(t)}}\! \nabla_{{\bm \theta}_{k}^{(t)}} {\psi}_{k} \left(\! {\bm \theta}_{k}^{(t)};\tilde{\bm \xi}_{k,i}^{(t)} \right)\nabla_{ {\psi}_{k} \left( {\bm \theta}_{k}^{(t)};\tilde{\bm \xi}_{k,i}^{(t)} \!\right) } \tilde{ \bm \psi}_{i}^{(t)} \nabla_{\tilde{ \bm \psi}_{i}^{(t)} } \!f_i\left(\!{\bm\theta}_0^{(t)};\tilde{\bm \psi}_{i}^{(t)} \!\right).\label{Sys_ChainRule}
\end{align}
\hrule
\end{figure*}
Note that the part $\nabla_{\tilde{ \bm \psi}_{i}^{(t)} }  f_i\left({\bm\theta}_0^{(t)};\tilde{\bm \psi}_{i}^{(t)} \right)$ could be sent from edge server to edge devices for back propagation, while the remaining execution $\nabla_{{\bm \theta}_{k}^{(t)}} {\psi}_{k} \left( {\bm \theta}_{k}^{(t)};\tilde{\bm \xi}_{k,i}^{(t)} \right) \nabla_{ {\psi}_{k} \left( {\bm \theta}_{k}^{(t)};\tilde{\bm \xi}_{k,i}^{(t)} \right) } \tilde{ \bm \psi}_{i}^{(t)}$ is executed locally. Then, both local models at edge devices and central server at the edge server would be updated by using gradient descent as 
\begin{align}\label{Sys-ModelUpdate}
	{\bm \theta}_{k}^{(t+1)}={\bm \theta}_{k}^{(t)}-\mu^{(t)} \hat{\bm g}\left({\bm \theta}_{k}^{(t)}\right), \forall k\in\{0\}\cup\mathcal{K},
\end{align}
where $\mu^{(t)}$ is the learning rate at the $t$-th round. 
\end{itemize}
The process repeats until the number of rounds $T$ is met.

\subsection{Sensing Model for Training Samples Acquisition}


Specifically, during the wireless sensing mode, each edge device transmits a dedicated \textit{frequency-modulated continuous wave} (FMCW) and then receives the corresponding echo signal, which serves as sensing data containing valuable information for training the AI models. It is assumed that all devices sense the same target from different perspectives, enabling them to obtain unique observations that provide diverse features for describing the target.


At any arbitrary round $t$, each device periodically transmits an FMCW signal with multiple up-chirps to illuminate the object, e.g., the human body.  Let $p_{k,\rm s}^{(t)}$ denote the sensing power at edge device $k$.
The received signal consists of three parts, including the desired normalized one-hop reflective signal, the clutter caused by multi-hop reflective paths, and the additive sensing noise.
By processing the corresponding echo signal, each edge device could collect a training data sample denoted by $\tilde{\bm \xi}_{k,i}^{(t)} $ through sampling, singular value decomposition, and short-time Fourier transform \cite{Liu-JSTSP2023,WenTWC25}, which is given by 
\begin{equation}\label{Sys:SensoryDataSample}
\tilde{\bm \xi}_{k,i}^{(t)}  ={ {\bm \xi}}_{k,i}^{(t)}+ {\bm \gamma}_{k}^{(t)} + \frac{{\bf n}_{s}^{(t)}  }{\sqrt{p_{k,s}^{(t)} }}, ~ \forall i\in \mathcal{B}_{k}^{(t)},\;  \forall k \in\mathcal{K},
\end{equation}
where ${ {\bm \xi}}_{k,i}^{(t)}$ is the ground-truth sample, ${\bm \gamma}_{k}^{(t)} $ is the clutter signal, ${\bf n}_s^{(t)}$ is the additive sensing noise following a zero-mean Gaussian distribution with  $\mathbb{E}\left(\left({{\bf n}_{s}^{(t)}}\right)^H{\bf n}_{s}^{(t)}\right) =\delta_s^2$. Without loss of generality, ${\bm \gamma}_{k}^{(t)}$ follows a zero-mean multi-variate Gaussian distribution with $\mathbb{E}=\left[\left({{\bm \gamma}_{k}^{(t)}}\right)^H{\bm \gamma}_{k}^{(t)}\right] = \delta_{k,s}^2$. 

Besides, due to the heterogeneity of sensing ability at different edge devices the latency for sensing one data sample is denoted as $\tau_{k, \rm s}^{(t)}$. The sensing time of device $k$ is given by 
\begin{align}\label{Sys_SensingTime}
T_{k, \rm s}^{(t)} = b^{(t)} \tau_{k, \rm s}^{(t)},~\forall k \in\mathcal{K},~\forall t\in\mathcal{T}.
\end{align}
Then, the sensing energy consumption of device $k$ is given by
\begin{align}
	E_{k,\rm s}^{(t)}=p_{k,s}^{(t)}T_{k, \rm s}^{(t)} = p_{k,s}^{(t)} b^{(t)} \tau_{k, \rm s}^{(t)},~\forall k \in\mathcal{K},~\forall t\in\mathcal{T}.
\end{align}

\subsection{Local Computation for V-FEEL}\label{ComputationModel}
At each round, edge devices would generate an embedding ${\psi}_{k} ( {\bm \theta}_{k}^{(t)};\tilde{\bm \xi}_{k,i}^{(t)} )$,  for each sample $\tilde{\bm \xi}_{k,i}^{(t)}, ~\forall i\in\mathcal{B}_{k}^{(t)}$ through pre-processing on the sensed data. 
 Let $C_k$ denote the {\it central processing unit} (CPU) cycles for execution of each sample and $\zeta_k$ denote the frequency.
 As there are a total of $b^{(t)}$ samples to be processed at each round, the computation latency is thus expressed as
 \begin{align}
 	T_{k,\rm c}^{(t)} =\frac{C_k b^{(t)}}{\zeta_k}, \forall k\in\mathcal{K},~\forall t\in\mathcal{T}.
 \end{align}
The energy consumption for local model updating at device $k$ in round $t$ is \cite{Cao19IoTJ}
\begin{align}
	E_{k,\rm c}^{(t)}= \kappa_k T_{k,\rm c}^{(t)} \zeta_k^3=\kappa_k C_k b^{(t)}\zeta_k^2,~\forall k \in\mathcal{K},~\forall t\in\mathcal{T},
\end{align}
where $\kappa_k$ represents the effective capacitance coefficient that depends on the chip architecture of edge device $k$ \cite{Burd96}.

\subsection{Embeddings Aggregation via Over-the-air Computation}\label{CommunicationModel}

In the VFEEL algorithm, only the aggregated local embeddings need to be uploaded to the edge server, eliminating the need to transmit local samples (features) or models. This significantly enhances privacy protection. Moreover, it also improves the communication efficiency, as the dimensionality of local embeddings is typically much lower than that of raw samples or model parameters. However, frequent communication for embedding uploads and aggregation would become a significant performance bottleneck, particularly when dealing with a large number of edge devices. 
To overcome this challenge, we leverage an AirComp approach, which enables the integration of communication and computation across multiple edge devices.


For ease of illustration, we employ a frequency non-selective block fading channel model, where wireless channels remain static within each global round but may vary across different rounds. Each edge device is assumed to have perfect knowledge of its own CSI, enabling phase pre-compensation at the transmitter. The edge server possesses global CSI to facilitate power control design. Let $\hat h_{k}^{(t)}$ denote the complex channel coefficient from edge device $k$ to the edge server at round $t$, and $h_{k}^{(t)}$ denote its magnitude with $h_{k}^{(t)}=|\hat h_{k}^{(t)}|$, $\forall k\in\mathcal{K}, t\in\mathcal{T}$.


With a minor abuse of notation, let ${\bm \psi}_{k} ( {\bm \theta}_{k}^{(t)};\tilde{\bm \xi}_{k}^{(t)} )\triangleq\left[ { \psi}_{k} \left( {\bm \theta}_{k}^{(t)};\tilde{\bm \xi}_{k,1}^{(t)} \right),\cdots,{ \psi}_{k} \left( {\bm \theta}_{k}^{(t)};\tilde{\bm \xi}_{k,b^{(t)}}^{(t)} \right)\right]$ denote the set of all embeddings associated with dataset $\mathcal{B}_{k}^{(t)}$ at edge device $k$ in round $t$. 
At any arbitrary round $t$, each edge device uses $q$ symbols to transmit embeddings when $b^{(t)}$ data samples are input into training, where $q = d b^{(t)}$ with $d$ denoting the dimensions of each embedding. 
Denote $p_{k}^{(t)}$ as the transmission power scaling factor. With proper phase control, edge devices are allowed to transmit simultaneously, and thus the received signal (after phase compensation) at the edge server is given by
 \begin{align}\label{Sys_ReceivedSignal}
 	{\bf y}^{(t)}=\sum\limits_{k\in\mathcal K}h_{k}^{(t)}\sqrt{p_{k}^{(t)}}{\bm \psi}_{k} \left( {\bm \theta}_{k}^{(t)};\tilde{\bm \xi}_{k}^{(t)} \right)+{\bm z}^{(t)}, \forall t\in\mathcal{T},
 \end{align}
 in which ${\bm z}^{(t)}$ denotes the {\it additive white Gaussian noise} (AWGN) with ${\bf z}^{(t)}\sim\mathcal{CN}(0,\sigma_z^2\bf I)$, as well as $\sigma_z^2$ and $\bf I$ are the noise power and an identity matrix, respectively.
 We also assume that each element of transmit signals ${\bm \psi}_{k} \left( {\bm \theta}_{k}^{(t)};\tilde{\bm \xi}_{k}^{(t)} \right)$ has zero mean and unit variance after normalization. 
 
Hence, the edge server estimates the global embeddings as $\tilde{ \bm \psi}^{(t)}$ by implementing a denoising factor $\eta^{(t)}$, i.e., 
\begin{align}\label{Sys_ComGlobalEmbed}
\tilde{ \bm \psi}^{(t)}&=\frac{{\bf y}^{(t)}}{\sqrt{\eta^{(t)}}}\notag\\
&=\frac{\sum\limits_{k\in\mathcal K}h_{k}^{(t)}\sqrt{p_{k}^{(t)}}{\bm \psi}_{k}^{(t)}\left( {\bm \theta}_{k}^{(t)};\tilde{\bm \xi}_{k}^{(t)} \right)+{\bf z}^{(t)}}{\sqrt{\eta^{(t)}}} , \forall t\in\mathcal{T}.
\end{align}

Recall that the size of transmitted parameters is denoted by $q$ and assume that each element of an embedding is modulated as a single analog symbol. To upload an embedding to the edge server, the total number of analog symbols to be transmitted is $q$.  Let $M$ denote the number of symbols in each resource block with duration $\tau_{\rm slot}$. 
At each round, the communication latency is thus expressed as 
\begin{align}\label{Sys_Latency_Air}
    T_{\rm t}=\left\lceil\frac{q}{M}\right\rceil \tau_{\rm slot}=\left\lceil\frac{d b^{(t)}}{M}\right\rceil \tau_{\rm slot},
\end{align}
where $\lceil \cdot\rceil$ denotes the integer ceiling function.
Notably, in LTE systems \cite{LTE}, each resource block within a duration of $T_{\rm slot}=1$ ms consists of two slots with $14$ symbols in general, and thus we have  $M=14$. 
Besides, the transmission energy consumption at each device is given by 
\begin{align}
	E_{k,\rm t}^{(t)}= p_{k}^{(t)}\tau_{\rm slot} ,~\forall k \in\mathcal{K},~\forall t\in\mathcal{T}.
\end{align}

\subsection{Network Resource Constraints}

During the training process, each edge device must operate under constraints related to latency, transmission power, and energy, due to limited network resources.

\subsubsection{Latency Constraints}

At each round, the latency includes three parts at each edge device in general, namely the data sensing, local computation, and embedding transmission. Meanwhile, the execution delay and model download time at the edge server are negligible, as the edge server always resides powerful base stations or access points with ample computational resources and energy supply.
Therefore, the total latency of each device should not exceed the allowed latency at each round denoted by $\Delta_{k}^{(t)},~\forall k\in\mathcal{K},~\forall t\in\mathcal{T}$, as given by
\begin{align}\label{Sys_Latency_Sum}
&T_{k, \rm s}^{(t)}+T_{k,\rm c}^{(t)} +T_{\rm t}\notag\\
&= b^{(t)} \tau_{k, \rm s}^{(t)}+\frac{C_k b^{(t)}}{\zeta_k}+\text{ceil}\left(\frac{d b^{(t)}}{M}\right)\tau_{\rm slot}
\leq\Delta_{k}^{(t)}.
\end{align}

\subsubsection{Transmission Power Constraints}
Due to the limited on-device battery, it is supposed that each device $k\in \mathcal K$ is subject to a maximum power budget $P^{\rm max}_k$ at each round $t$, as given by 
\begin{align}
&\frac{1}{d b^{(t)}}\!\mathbb{E}\!\left[p_{k}^{(t)}\left\|\!{\bm \psi}_{k} \left( {\bm \theta}_{k}^{(t)};\tilde{\bm \xi}_{k}^{(t)} \!\right)\!\right\|^2\!\right]\!\!=\!\! \frac{p_{k}^{(t)} }{d b^{(t)}}  \!\leq \!P^{\rm max}_k,\label{sys_bar_P_max1}\\
  &p_{k,s}^{(t)}  \leq P^{\rm max}_k,~\forall k\in{\mathcal K}, ~ t\in\mathcal{T}.\label{sys_bar_P_s_max1}
\end{align}

\subsubsection{Energy Consumption Constraints}

At each round,  the energy consumption of edge device $k$ also consists of three parts for the sensing, local computation, and concurrent aggregation via AirComp. 
We thus have the following constraint for the $k-$th edge device across all rounds.
\begin{align}
\!\!\!\!&\!\sum\limits_{t\in\mathcal{T}}\left(E_{k,\rm s}^{(t)} +E_{k,\rm c}^{(t)}+E_{k,\rm t}^{(t)}\right)=\notag\\
\!\!\!\!&\!\sum\limits_{t\in\mathcal{T}}\!\!\left(p_{k,s}^{(t)} b^{(t)} \tau_{k, \rm s}^{(t)}\!+\! \kappa_k C_k b^{(t)}\!\zeta_k^2\!+\!p_{k}^{(t)}\!\tau_{\rm slot} \!\right) \! \leq E_k,\forall k\in{\mathcal K}.\label{Sys_Energy}
\end{align}

\section{Convergence Analysis of ISCC-based V-FEEL}\label{Sect:Convergence}

In this section, we present a convergence analysis for the ISCC-enabled V-FEEL, accounting for both sensing and aggregation errors.

\subsection{Basic Assumptions}

For the purpose of analysis,  we first adopt several assumptions on the loss function and embedding estimates as follows, which have been commonly adopted in the literature \cite{Castiglia22PMLR,Tran2025PBM}. 

 \begin{assumption}[Smoothness]\label{Assump_Smooth}\emph{
The gradient of loss functions is Lipschitz continuous with a common non-negative constant $L>0$. And for any ${\bm \Theta}_1, {\bm \Theta}_2 \in\mathbb{R}^{V}$, it holds that 
\begin{align}
\left\|\nabla	 F({\bm \Theta}_1)-\nabla F({\bm \Theta}_2)\right\| & \leq L\|{\bm \Theta}_1- {\bm \Theta}_2\|,\label{Smooth_gra_k}
\end{align}
where $\nabla F(\cdot )$ denotes the gradients of the loss function evaluated at points ${\bm \Theta}$. 
As a consequence, for any ${\bm \Theta}_1, {\bm \Theta}_2 \in\mathbb{R}^{V}$, we have 
\begin{align}
F({\bm \Theta}_1) &\le  F({\bm \Theta}_2) + \nabla F({\bm \Theta}_2)^T ({\bm \Theta}_1-{\bm \Theta}_2)+\frac{L}{2}\|{\bm \Theta}_1-{\bm \Theta}_2\|^2. \label{smooth_Func}
\end{align}
}
\end{assumption}

Let ${\bm g}_i\left({\bm \theta}_{k}^{(t)}\right)$ and ${\bm \psi}_{i}^{(t)}=\sum\limits_{k\in\mathcal{K}}  {\psi}_k ({\bm \theta}_k;{\bm \xi}_{k,i})$ denote gradient and aggregated embeddings over the clean data sample ${\bm \xi}_{k,i}^{(t)}$ with error-free data aggregation, respectively. It thus holds the following chain rule of gradients execution.
\begin{align*}
	&{\bm g}_i\left({\bm \theta}_{0}^{(t)}\right)\triangleq \nabla_{0}f_i\left({\bm\theta}_0^{(t)};{\bm \psi}_{i}^{(t)} \right)\notag\\
    &=\nabla_{{\bm \theta}_{0}^{(t)}}{ \psi}_0^{(t)}\left({\bm\theta}_0^{(t)};{\bm \psi}_{i}^{(t)} \right)\nabla_{{ \psi}_0^{(t)}\left({\bm\theta}_0^{(t)};{\bm \psi}_{i}^{(t)} \right) }  f_i\left({\bm\theta}_0^{(t)};{\bm \psi}_{i}^{(t)} \right);\\
	&{\bm g}_i\left({\bm \theta}_{k}^{(t)}\right)\triangleq \nabla_{k}f_i\left({\bm\theta}_0^{(t)};{\bm \psi}_{i}^{(t)} \right)\notag\\
    &= \nabla_{{\bm \theta}_{k}^{(t)}} {\psi}_{k} ( {\bm \theta}_{k}^{(t)};{\bm \xi}_{k,i}^{(t)} ) \nabla_{ {\psi}_{k} ( {\bm \theta}_{k}^{(t)};{\bm \xi}_{k,i}^{(t)} ) } { \bm \psi}_{i}^{(t)} \nabla_{{ \bm \psi}_{i}^{(t)} }  f_i\left({\bm\theta}_0^{(t)}; { \bm \psi}_i^{(t)} \right),
\end{align*}
based on which, ${\bm g}\left({\bm \theta}_{k}^{(t)}\right)=\frac{1}{b^{(t)}}\sum_{i=1}^{b^{(t)}} {\bm g}_i\left({\bm \theta}_{k}^{(t)}\right),\forall k\in\{0\}\cup\mathcal{K}$.
Without loss of generality, we also made the following common assumptions on gradients \cite{Bottou2018,Tsitsiklis1986TAC}.

 \begin{assumption}[Unbiased Gradient with Bounded Variance]\label{Assum_VarianceBound}
 \emph{The gradient is unbiased with bounded variance, given by 
 \begin{align*}
&\mathbb{E}\left[ {\bm g}_i\left({\bm \theta}_{k}^{(t)}\right)\right]=\nabla_k F({\bm \Theta}^{(t)} ), \forall i\in\mathcal{B}_{k}^{(t)},k\in\{0\}\cup\mathcal{K}, t\in\mathcal{T};\\
&\mathbb{E}\left\|{\bm g}_i\!\!\left(\!{\bm \theta}_{k}^{(t)}\right)\!\!-\! \nabla_k F({\bm \Theta}^{(t)} )\! \right\|^{2} \!\!\! \leq \sigma^{2},\!\forall i\in\mathcal{B}_{k}^{(t)}\!,k\in\{0\}\cup\mathcal{K},\! t\in\mathcal{T},\!
  \end{align*}
 where $\nabla_k F\left({\bm \Theta}^{(t)}\right)$ represents the ground-truth gradient for $k\in\{0\}\cup\mathcal{K}$ at round $t$, $\mathbb{E}(\cdot)$ denotes the statistical expectation, and the non-negative constant $\sigma^{2}$ is the per sample gradient variance.}
\end{assumption}

Following \cite{Castiglia22PMLR}, the following assumptions of bounded norm is made on the embedding functions.
\begin{assumption}[Bounded Hessian]\label{Assump_BoundedHessian} \emph{There exist positive constants $\Psi_k$ for $k\in\mathcal{K}$ such that for all samples, the second partial derivatives of sample-wise loss function w.r.t. local embedding ${\psi}_{k} \left( {\bm \theta}_{k}^{(t)};{\bm \xi}_{k,i}^{(t)} \right) $ satisfy:
\begin{align}
   \left\| \nabla_{ {\psi}_{k}}^2 f_i\left({\bm\theta}_0^{(t)};{\bm \psi}_{i}^{(t)} \right)\right\|_{\mathcal{F}}\leq \Psi, \forall k\in\mathcal{K},~1\leq  i\leq b^{(t)},
\end{align}
where ${\mathcal{F}}$ denotes the Frobenius norm.}
\end{assumption}

\begin{assumption}[Bounded Embedding Gradients]
\label{Assump_BoundedEmbedGra} \emph{There also exist positive constants $G_1$ and $G_2$ such that for model parameter $\theta_k^{(t)}$ and data sample ${\bm \xi}_{k,i}^{(t)}$, respectively, the partial embedding gradients are bounded by 
\begin{align*}
   \left\| \nabla_{{\bm \theta}_{k}^{(t)}} {\psi}_{k} \left( {\bm \theta}_{k}^{(t)};{\bm \xi}_{k,i}^{(t)} \right)   \right\|_{\mathcal{F}}\leq G_1, \forall k\in\mathcal{K},~1\leq  i\leq b^{(t)}, ~t\in\mathcal{T};\\
      \left\| \nabla_{{\bm \xi}_{k,i}^{(t)}} {\psi}_{k} \left( {\bm \theta}_{k}^{(t)};{\bm \xi}_{k,i}^{(t)} \right)   \right\|_{\mathcal{F}}\leq G_2, \forall k\in\mathcal{K},~1\leq  i\leq b^{(t)}, ~t\in\mathcal{T}.
\end{align*}}
\end{assumption}

Note that at each round $t$, each edge device $k$ trains its local model under the sensed noisy data ${\bm \xi}_{k,i}^{(t)}$ in \eqref{Sys:SensoryDataSample}, and then outputs its ground-truth embedding ${ \psi}_{k} ( {\bm \theta}_{k}^{(t)};{\bm \xi}_{k,i}^{(t)} )$, $\forall i\in\mathcal{B}_{k}^{(t)}$.  
 However, AirComp-induced aggregation errors corrupt the received embeddings at the edge server. This corruption leads to an erroneous gradient in the model update \eqref{Sys-ModelUpdate}. Consequently, both two distinct errors affect the aggregated signal in \eqref{Sys_ComGlobalEmbed}: namely the inherent data noise ${\bm \xi}_{k,i}^{(t)}$ and the AirComp aggregation error.
In the following, we first quantify the impact of sensory data noise on the loss function and then establish the convergence properties in the presence of aggregation error.

\subsection{Data and Aggregation Error Analysis}

Recall the local embedding vector ${ \psi}_{k} \left( {\bm \theta}_{k}^{(t)};\tilde{\bm \xi}_{k,i}^{(t)} \right)$ at each edge device  is generated with a noisy data sample $ \tilde{\bf \xi}_{k,i}^{(t)}$ defined in \eqref{Sys:SensoryDataSample}. 
As for each clean data sample ${\bf \xi}_{k,i}^{(t)}$, the corresponding local embedding is denoted by ${ \psi}_{k} \left( {\bm \theta}_{k}^{(t)};{\bm \xi}_{k,i}^{(t)} \right)$. 
By taking the Taylor expansion of the embedding function with noisy data at the reference point ${\bm \xi}_{k,i}^{(t)}$ is given by
\begin{align}\label{Ana_EmbeddingNoise}
	 &{ \psi}_{k} \left( {\bm \theta}_{k}^{(t)};\tilde{\bm \xi}_{k,i}^{(t)} \right) =  \nabla_{{\bm \xi}_{k,i}^{(t)}} { \psi}_{k} \left( {\bm \theta}_{k}^{(t)};{\bm \xi}_{k,i}^{(t)} \right)\left(\tilde{\bm \xi}_{k,i}^{(t)} - {\bm \xi}_{k,i}^{(t)}\right)\notag\\ 
     &~~~~~~~~~~~~~~~~~~~~+{ \psi}_{k} \left( {\bm \theta}_{k}^{(t)};{\bm \xi}_{k,i}^{(t)} \right)+ O\left(\tilde{\bm \xi}_{k,i}^{(t)} - {\bm \xi}_{k,i}^{(t)}\right),
\end{align}
where $O\left(\tilde{\bm \xi}_{k,i}^{(t)} - {\bm \xi}_{k,i}^{(t)}\right)$ is the infinitesimal of higher order, which could be ignored \cite{Castiglia22PMLR} due to the fact that . 

Next, we substitute the sensing data sample noise in \eqref{Sys:SensoryDataSample} into \eqref{Ana_EmbeddingNoise} and further ignore the infinitesimal of higher order terms, and it holds 
\begin{align}\label{Eq:TaylorApproximation}
&{ \psi}_{k} \left( {\bm \theta}_{k}^{(t)};\tilde{\bm \xi}_{k,i}^{(t)} \right) \approx   \nabla_{{\bm \xi}_{k,i}^{(t)}} { \psi}_{k} \left( {\bm \theta}_{k}^{(t)};{\bm \xi}_{k,i}^{(t)} \right) \left({\bm \gamma}_k^{(t)} + \dfrac{  {\bf n}_s^{(t)}}{\sqrt{p_{k,s}^{(t)}} } \right)\notag\\
&~~~~~~~~~~~~~~~~~~~~+ { \psi}_{k} \left( {\bm \theta}_{k}^{(t)};{\bm \xi}_{k,i}^{(t)} \right).
\end{align}
Let $\! {\bm \psi}_{k} \left( {\bm \theta}_{k}^{(t)};{\bm \xi}_{k}^{(t)} \right)\!=\!\left[ { \psi}_{k} \left( {\bm \theta}_{k}^{(t)};{\bm \xi}_{k,1}^{(t)} \right),\!\cdots,\!  { \psi}_{k} \left( {\bm \theta}_{k}^{(t)};{\bm \xi}_{k,b^{(t)}}^{(t)} \right)\right]  $ define the set of ground-truth embeddings at edge device $k$. 
Let  ${ \bm \psi}_{i}^{(t)}$ be the desired information at edge server for each sample $i$ is ${\bm \psi}_{i}^{(t)} \triangleq\sum \limits_{k\in\mathcal{K}} { \psi}_{k} \left( {\bm \theta}_{k}^{(t)};{\bm \xi}_{k,i}^{(t)} \right).$
Therefore, based on \eqref{Sys_ComGlobalEmbed}, the aggregation error caused by the AirComp w.r.t. the global embedding estimation $\tilde{ \bm \psi}^{(t)}_{i}$ at $i$-th sample is given by
\begin{align}\label{Sys_Err}
&{\bm \varepsilon}_{i}^{(t)}=\tilde{ \bm \psi}_{i}^{(t)}-{ \bm \psi}_{i}^{(t)}\notag\\
&=\!\sum\limits_{k\in\mathcal K}\!\left(\!\!\frac{h_{k}^{(t)}\sqrt{p_{k}^{(t)}}}{\sqrt{\eta^{(t)}}} { \bm \psi}_{k}^{(t)}\left( {\bm \theta}_{k}^{(t)};\tilde{\bm \xi}_{k,i}^{(t)} \right)\! -\! { \bm\psi}_{k}^{(t)}\!\!\left(\! {\bm \theta}_{k}^{(t)};{\bm \xi}_{k,i}^{(t)} \!\right) \!\right)\!+\!\frac{{ z}_{i}^{(t)}}{\sqrt{\eta^{(t)}}} \notag\\
&=\sum\limits_{k\in\mathcal K}\left(\frac{h_{k}^{(t)}\sqrt{p_{k}^{(t)}}}{\sqrt{\eta^{(t)}}} -1 \right){\bm \psi}_{k} \left( {\bm \theta}_{k}^{(t)};{\bm \xi}_{k,i}^{(t)} \right)\notag\\
&~~+\sum\limits_{k\in\mathcal K}\frac{h_{k}^{(t)}\sqrt{p_{k}^{(t)}}}{\sqrt{\eta^{(t)}}}\left({\bm \gamma}_k^{(t)} + \dfrac{  {\bf n}_s^{(t)}}{\sqrt{p_{k,s}^{(t)}} } \right) \nabla_{{\bm \xi}_{k,i}^{(t)}} { \psi}_{k} \left( {\bm \theta}_{k}^{(t)};{\bm \xi}_{k,i}^{(t)} \right)\notag\\
&~~+\frac{{ z}_{i}^{(t)}}{\sqrt{\eta^{(t)}}}, ~1\leq  i\leq b^{(t)},
\end{align}
where the subscript character $i$ denotes the $i$-th sample. 
Note that from \eqref{Sys_Err}, it has $\tilde{ \bm \psi}^{(t)}_i={\bm \varepsilon}_i^{(t)}+{ \bm \psi}_i^{(t)}$, which indicates that the obtained gradient in \eqref{Sys_ChainRule} is erroneous.
In this case, at each round $t$, the statistical property (e.g. bias and MSE) of embedding estimates through the over-the-air aggregation is derived as
\begin{align}
	&\mathbb{E}\left[ {\bm \varepsilon}_{i}^{(t)} \right]=0; \label{Ana_AggBias}\\
	&\mathbb{E} \!\left\|\!{\bm \varepsilon}_{i}^{(t)}\! \right\|^2\!\!\! 
	\leq\! \sum\limits_{k\in\mathcal K}\!\left(\!\frac{h_{k}^{(t)}\sqrt{p_{k}^{(t)}}}{\sqrt{\eta^{(t)}}} \!-1 \!\right)^2 \!\!\mathbb{E} \left\|{\bm \psi}_{k} \left( {\bm \theta}_{k}^{(t)};{\bm \xi}_{k,i}^{(t)} \right) \!\right\|^2\!\!+\!\frac{\sigma_z^2}{\eta^{(t)}}\notag\\
	&+\!\!\sum\limits_{k\in\mathcal K}\!\!\frac{\left(h_{k}^{(t)}\right)^2p_{k}^{(t)}}{\eta^{(t)}}\mathbb{E}\!\! \left\|\!\left(\!{\bm \gamma}_k^{(t)}\! \!+\! \frac{ \! {\bf n}_s^{(t)}}{\sqrt{p_{k,s}^{(t)}} } \!\right)\! \!\nabla_{{\bm \xi}_{k,i}^{(t)}} { \psi}_{k} \left(\! {\bm \theta}_{k}^{(t)};{\bm \xi}_{k,i}^{(t)} \!\right)\!\right\|^2\!\label{Ana_AggMSE2}\\
	&\!\!\leq\! \!\sum\limits_{k\in\mathcal K}\!\!\left(\!\frac{h_{k}^{(t)}\sqrt{p_{k}^{(t)}}}{\sqrt{\eta^{(t)}}}\! -\!1 \!\!\right)^2\!\!\!\!+\!\!\sum\limits_{k\in\mathcal K}\!\frac{\left(h_{k}^{(t)}\!\right)\!^2p_{k}^{(t)}}{\eta^{(t)}}\!\left(\! \delta_{k,s}^2\!+\!\frac{\delta_s^2}{p_{k,s}^{(t)}}\!\right)\!\!G_2^2\!+\!\frac{\sigma_z^2}{\eta^{(t)}}\!\notag\\
    &\triangleq \mathbb{E} \left\|\tilde{\bm \varepsilon}^{(t)} \right\|^2\label{Ana_AggMSE3},
\end{align}
where \eqref{Ana_AggMSE2} holds since the sensing noise and channel noise have zero means, \eqref{Ana_AggMSE3} holds due to Assumption \ref{Assump_BoundedEmbedGra}, and $\mathbb{E}\left[ \left\|\tilde{\bm \varepsilon}^{(t)} \right\|^2\right]$ represents the MSE bound of aggregation error.

\begin{remark}\emph{
     According to \eqref{Ana_AggMSE3}, the MSE of aggregation error consists of three parts including the alignment error, wireless sensing-induced sample noise, and transmission noise-induced error, all of which are balanced by the receive denoising factor $\eta^{(t)}$. Specifically, raising the denoising factor $\eta^{(t)}$ can significantly diminish the sensing and noise-induced error at the expense of increased misalignment error. 
     Moreover, increasing sensing power can suppress sensing noise, while concurrently reduces the available transmission power as constrained by \eqref{sys_bar_P_max1}. This reduction in transmission power leads to a higher misalignment error.
	Note that it is also observed that it is independent of the sensing data samples but related to the sensing strategy (e.g., sensing power, batch size of sensing sample, and variances of clutter and sensing noise). In other words, this decoupling simplifies system design, as the joint consideration of sensing approach and embedding aggregation dictates MSE performance.}
\end{remark}

To account for aggregation error, using the chain rule and Taylor series expansion, we have the following lemma to bound the difference between $\hat{\bm g}\left({\bm \Theta}^{(t)}\right)$ and $\nabla F\left({\bm \Theta}^{(t)}\right)$.

\begin{lemma}[Unbiased and Bounded Embedding Gradient Vector]\label{Lemma_Embed}\emph{At each round $t$, for edge device $k\in\mathcal{K}$, the partial derivative of  is unbiased:
\begin{align}
	\mathbb{E}\left( 	\hat{\bm g}\left({\bm \theta}_{k}^{(t)}\right)\right)&=\nabla_k F({\bm \Theta}^{(t)} ),\forall k\in\{0\}\cup\mathcal{K}, t\in\mathcal{T},
\end{align}
Then the variances of the partial derivatives are bounded as
\begin{align}
	&\mathbb{E}\left[ \left\|	\hat{\bm g}\left({\bm \theta}_{k}^{(t)}\right)-\nabla_k F({\bm \Theta}^{(t)} )\right\|^2\right]\leq \notag\\
    &\frac{\sigma^2}{b^{(t)}}+\frac{G_1^2 \Psi^2}{b^{(t)}} \mathbb{E}\left[\left\|\tilde{\bm \varepsilon}^{(t)}\right\|^2\right], \forall k\in\{0\}\cup\mathcal{K},~t\in\mathcal{T}.
\end{align}
	}
\end{lemma}


\begin{IEEEproof}
Please refer to Appendix \ref{Theo_Lemma_Embed}.
\end{IEEEproof}

\subsection{Convergence Analysis}

In this subsection, we discuss the convergence behavior of the proposed ISCC-enabled V-FEEL algorithm by investigating the loss function descent.

Based on Assumptions \ref{Assump_Smooth}-\ref{Assump_BoundedEmbedGra} and Lemma \ref{Lemma_Embed}, we could bound the per-round loss function gap by considering a properly chosen fixed learning rate, which can be derived as follows.
\begin{lemma}[Per-Round Loss Function Reduction]\label{Lemma_Loss_Reduction}\emph{With any given $0\leq \mu^{(t)}< \frac{2}{L}, \forall t\in\mathcal{T}$, the expected per-round loss descent satisfies 
\begin{align}
	&\mathbb{E}\left[F\left({\bm \Theta}^{(t+1)}\right)- F\left({\bm \Theta}^{(t)}\right) \right]\notag\\
&\leq - \mu^{(t)}\left(1-\frac{L(\mu^{(t)})^2}{2}\right) \left\|\nabla F\left({\bm \Theta}^{(t)}\right) \right\|^{2}  \notag\\
&~~+\frac{L(\mu^{(t)})^2\left(K+1\right)(\sigma^2+G_1^2 \Psi^2)}{2b^{(t)}} \mathbb{E}\left[\left\|\tilde{\bm \varepsilon}^{(t)}\right\|^2\right]\label{Lemma_Loss}.
\end{align}	
}
\end{lemma}

\begin{IEEEproof}
Please refer to Appendix \ref{Lemma_Loss_Reduction_Proof}.
\end{IEEEproof}

\begin{remark}\emph{
Lemma \ref{Lemma_Loss_Reduction} reveals that a larger batch size $b^{(t)}$ accelerates the empirical loss descent per round, yet this acceleration is impeded by aggregation error $\mathbb{E}\left[\left\|\tilde{\bm \varepsilon}^{(t)}\right\|^2\right]$.
Coupled with the insight from \eqref{Ana_AggMSE3}, effective reduction of this error necessitates careful design of the denoising factor $\eta^{(t)}$. 
We observe that increasing the number of edge devices (i.e., $K+1$) amplifies the impact of aggregation error on convergence rate. This occurs because all edge devices and edge server incorporate the induced aggregation error into their local model updates in \eqref{Sys-ModelUpdate}, thereby degrading model updates and slowing convergence. This stands in sharp contrast to horizontal AirFEEL system, where involving more devices typically could enhance learning performance \cite{WenTWC25,Cao2021-FedAvg}.   }
\end{remark}
 

\begin{theorem}
	[Convergence with Fixed Learning Rate]\label{Theo_Gradient}\emph{
Considering a fixed learning rate $\mu=\mu^{(t)}, \forall t\in\mathcal{T}$ with $0\leq \mu<\frac{2}{L}$, the expected loss descent under any given number of communication rounds $T$ satisfies 
\begin{align}\label{Theo_Gradient_Eq}  
&\frac{1}{T}\sum_{t=1}^{T}\left\|\nabla F\left({\bm \Theta}^{(t)}\right) \right\|^{2} \leq \frac{2\mathbb{E}\left[F\left({\bm \Theta}^{(1)}\right)- F\left({\bm \Theta}^{(T+1)}\right) \right]}{\mu\left(2-L\mu\right)T} \notag\\
&~~~~~~~~~~~~~~~~~~~~~~~~+\Omega\left( \left\{p_{k}^{(t)},p_{k,s}^{(t)},b^{(t)},\eta^{(t)}\right\} \right),
\end{align}
where $\Omega\left( \left\{p_{k}^{(t)},p_{k,s}^{(t)},b^{(t)},\eta^{(t)}\right\} \right)$ is defined as in \eqref{Sys_Omega}.
}
with $c_1= \frac{L\mu(K+1)(\sigma^2+G_1^2 \Psi^2)}{\left(2-L\mu\right)T}$.
\end{theorem}
\begin{IEEEproof}
Please refer to Appendix \ref{Theo_Gradient_Proof}.
\end{IEEEproof}

\begin{figure*}
\begin{align}
&\Omega\left( \left\{p_{k}^{(t)},p_{k,s}^{(t)},b^{(t)},\eta^{(t)}\right\} \right)\!\triangleq\!\sum_{t=1}^{T}\frac{c_1(K+1)}{b^{(t)}} \left[ \sum\limits_{k\in\mathcal K}\left(\frac{h_{k}^{(t)}\sqrt{p_{k}^{(t)}}}{\sqrt{\eta^{(t)}}} -1 \right)^2+\sum\limits_{k\in\mathcal K}\frac{\left(h_{k}^{(t)}\right)^2p_{k}^{(t)}}{\eta^{(t)}}\left( \delta_{k,s}^2+\frac{\delta_s^2}{p_{k,s}^{(t)}}\right)G_2^2+\frac{\sigma_z^2}{\eta^{(t)}}\right],\!\label{Sys_Omega}
\end{align}	
\hrule
\end{figure*}

\begin{remark}\label{Remark_leanding}
\emph{
	The first term in the bound in \eqref{Theo_Gradient_Eq}  represents the optimization gap between the initial loss and the loss after $T$ rounds, while it vanishes as $T\to\infty$. The second part is the aggregation error associated with variance of the sensing data, the number of edge devices, and the Lipschitz constant $L$. Crucially, this error scales inversely with the batch size $b^{(t)}$ and thus diminishes as $b^{(t)}$ increases.  However, under constrained network resources, a larger batch size requires more time or power for data sensing, which increases aggregation error and further degrades learning performance.  
    As established in Remark 1, joint transmission and sensing power control is crucial for minimizing the aggregation error $\mathbb{E}\left[\left\|\tilde{\bm \varepsilon}^{(t)}\right\|^2\right]$. This thus highlights the necessity for joint optimization of batch size and power allocation.
	}
\end{remark}


\section{Joint Batch Size and Power Allocation Optimization}\label{Sec_Opt}

Given the convergence results of ISCC-enabled V-FEEL in the preceding section, this section is ready to present the joint optimization of batch size and power control polices for accelerating the convergence.

\subsection{Problem Formulation}
With the obtained Theorem~\ref{Theo_Gradient}, we aim to speed up the convergence rate by minimizing the dominated term $\Omega\left( \left\{p_{k}^{(t)},p_{k,s}^{(t)},b^{(t)},\eta^{(t)}\right\} \right)$, namely the second term in \eqref{Theo_Gradient_Eq}, which is related to the inversion of the batch size $b^{{(t)}}$ and aggregation error $ \mathbb{E}\left\|\tilde{\bm \varepsilon}^{(t)}\right\|^2$. Therefore, the optimization problem is formulated as 
\begin{align}
\textrm{(P1):}~~\min\limits_{\{p_{k}^{(t)}\geq 0,p_{k,s}^{(t)}\geq 0,b^{(t)}\in \mathbb{Z}^{+},\eta^{(t)}\geq 0\}} & \Omega\left( \left\{p_{k}^{(t)},p_{k,s}^{(t)},b^{(t)},\eta^{(t)}\right\} \right)\notag\\
\textrm{s.t.} ~~~~~~~~~~~~~~~& \eqref {Sys_Latency_Sum}- \eqref{Sys_Energy},\notag
\end{align}
where $ \mathbb{Z}^{+}$ denotes the set of positive integers.
Note due to the coupling among sensing power, transmission power, and denoising factor, the primary problem is a non-convex and mixed-integer problem, which is hard to be tackled.

To deal with this difficulty, we adopt the alternating optimization technique to problem (P1), which are divided into two sub-problems, as described in the following.


\subsection{Optimization of Sensing Power and Size of Data Batch}

Under any given transmission power and denoising factor, the primary optimization problem (P1) is reduced into 
\begin{align}
\textrm{(P1.1):}\!\!\!\min\limits_{\{p_{k,s}^{(t)}\geq 0,b^{(t)}\in \mathbb{Z}^{+}\}} & 
 \sum_{t=1}^{T}\!\frac{\tilde{c}_1}{b^{(t)}}\! \left[ \!A_1^{(t)}+G_2^2\!\sum\limits_{k\in\mathcal K}\!\!\frac{\left(h_{k}^{(t)}\right)^2p_{k}^{(t)}\delta_s^2}{\eta^{(t)}p_{k,s}^{(t)}}\right]\notag\\
\textrm{s.t.} ~~~~~~~& \eqref {Sys_Latency_Sum}-\eqref{Sys_Energy}
\notag,
\end{align}
where $\tilde{c}_1=c_1(K+1) $ and $A_1^{(t)}=\sum\limits_{k\in\mathcal K}\left(\frac{h_{k}^{(t)}\sqrt{p_{k}^{(t)}}}{\sqrt{\eta^{(t)}}} -1 \right)^2+G_2^2\sum\limits_{k\in\mathcal K}\frac{\left(h_{k}^{(t)}\right)^2p_{k}^{(t)}\delta_{k,s}^2}{\eta^{(t)}}+\frac{\sigma_z^2}{\eta^{(t)}}$. Note that problem (P1.1) is still non-convex due to the coupling of  sensing power $\{p_{k,s}^{(t)}\}$ and the size of data batch $b^{(t)}$ as well as the integrity of $b^{(t)}$. To optimally solve this problem, we introduce a series of auxiliary variables as $e_{k,\rm s}^{(t)}=p_{k,\rm s}^{(t)}b^{(t)}, \forall k\in\mathcal{K},~t\in\mathcal{T}$ and relaxing the integer $b^{(t)}\in \mathbb{Z}^{+}$ into a continuous variable as $b^{(t)}\geq 0$, $\forall t\in\mathcal{T}$. 
 By recasting the objective function and ignoring the maximum sensing power constraints in \eqref{sys_bar_P_s_max1}, problem (P1.1) is thus reformulated as 
\begin{align}
&\min\limits_{\left\{e_{k,s}^{(t)}\geq 0,b^{(t)}\geq 0\right\}}  
  \sum_{t=1}^{T}\!\left[\!\frac{A_1^{(t)}\tilde{c}_1}{b^{(t)}} \!+ \!\sum\limits_{k\in\mathcal K}\!\frac{ \tilde{c}_1G_2^2\left(h_{k}^{(t)}\right)^2p_{k}^{(t)}\delta_s^2}{\eta^{(t)}e_{k,s}^{(t)}} \!\right]\label{SubP_sensing2}\\
&~~~~~~~~\textrm{s.t.} ~~\!\sum\limits_{t\in\mathcal{T}}\!\left(\!e_{k,s}^{(t)} \tau_{k, \rm s}^{(t)}\!\!+ \!\!\kappa_k C_k b^{(t)}\zeta_k^2\!+\!p_{k}^{(t)}\tau_{\rm slot} \!\right) \! \! \leq \!E_k,\!\forall k\in{\mathcal K} \label{SubP_sensing2C1}\\
&~~~~~~~~~~~~\!b^{(t)}\!\! \left(\!\!\tau_{k, \rm s}^{(t)}\!+\!\frac{C_k }{\zeta_k}\!+\!\frac{d \tau_{\rm slot}}{M}\! \right)\!
\leq\!\Delta_{k}^{(t)}\!,\forall k\in\mathcal{K}, t\in\mathcal{T}\!\label{SubP_sensing2C2}\\ 
&~~~~~~~~~~~~b^{(t)}\! \geq\! \frac{p_{k}^{(t)}}{d P^{\rm max}_k},~\forall k\in{\mathcal K},~ t\in\mathcal{T}\label{SubP_sensing2C3},
\end{align}
where the total latency constraint in \eqref{SubP_sensing2C2} is a relaxed counterpart of \eqref{Sys_Latency_Sum} and constraint \eqref{SubP_sensing2C3}
is from \eqref{sys_bar_P_max1}. Thus, problem \eqref{SubP_sensing2} is a convex problem w.r.t. $e_{k,s}^{(t)}$ and $b^{(t)}$. 

By leveraging the Lagrange duality method, we have the following proposition.

\begin{proposition}\label{SubP_sensing2_Prop}\emph{ 
With any given transmission power and denoising factor, the optimal solution to problem \eqref{SubP_sensing2} denoted by $\left\{e_{k,s}^{(t)\star}\right\}$ and $\left\{b^{(t)\star}\right\}$  is given by 
\begin{align}
\!\!e_{k,s}^{(t)\star}&\!=\!\sqrt{ \frac{ \tilde{c}_1 G_2^2  \left(h_{k}^{(t)}\right)^2p_{k}^{(t)}\delta_s^2 }{\lambda_k^{\star}\tau_{k, \rm s}^{(t)} \eta^{(t)} } }, ~\forall k\in{\mathcal K}, t\in\mathcal{T},\!\label{Prop1_e}\\
b^{(t)\star}&= \left(\sqrt{ \frac{\tilde{c}_1 A_1^{(t)} }{ \sum \limits_{k\in\mathcal{K}}\lambda_k^{\star} \kappa_k C_k \zeta_k^2  }}    \right)_{b^{(t),\rm l}}^{b^{(t),\rm u}},~\forall t\in\mathcal{T},\label{Prop1_b}
\end{align}
where $\lambda_k^{\star}$ is the optimal dual variable associated with the $k$-th energy constraint in \eqref{SubP_sensing2C1}, 
 $( x)_{u_1}^{u_2}\triangleq\min(u_2,\max(u_1,x))$ with $b^{(t),\rm l}=\max \limits_{k\in\mathcal{K}}\frac{p_{k}^{(t)} }{d P^{\rm max}_k}$ and $b^{(t),\rm u}=\min \limits_{k\in\mathcal{K}} \tilde\Delta_{k}^{(t)}$, and $\tilde\Delta_{k}^{(t)}\triangleq \frac{\Delta_{k}^{(t)}}{ \left(\tau_{k, \rm s}^{(t)}+\frac{C_k }{\zeta_k}+\frac{d \tau_{\rm slot}}{M} \right)}, ~\forall k\in\mathcal{K},~\forall t\in\mathcal{T}$.
	}
\end{proposition}

\begin{IEEEproof}
See Appendix~\ref{Proof_SubP_sensing2_Prop}.
\end{IEEEproof}

With obtained optimal $e_{k,s}^{(t)\star}$ in \eqref{Prop1_e}, we need to further construct the optimal $p_{k,\rm s}^{(t)},~\forall k\in{\mathcal K}, ~ t\in\mathcal{T}$ as given by 
\begin{align}\label{SubP_sensing2_Pks}
	&p_{k,\rm s}^{(t)\star}=\min\left(\frac{e_{k,\rm s}^{(t)}\star}{b^{(t)\star} }, P^{\rm max}_k\right)\notag\\
    &=\min\left(\sqrt{ \frac{\tilde{c}_1 G_2^2  \left(h_{k}^{(t)}\right)^2p_{k}^{(t)}\delta_s^2 }{\lambda_k^{\star}\tau_{k, \rm s}^{(t)} \eta^{(t)} (b^{(t)\star})^2} }, P^{\rm max}_k\right).
\end{align} 
Furthermore, we proceed to reconstruct an optimal solution to problem (P1.1). By rounding the optimal $b^{(t)\star}$ in \eqref{Prop1_b} into integers, the optimal size of data batch in problem (P1.1) is obtained, This integer solution is then substituted back into (P1.1) to optimize the sensing power, thereby ensuring that the resource constraints are not violated.

\begin{remark}\emph{Notice that the size of bath size is constrained by a region, where the upper bound and lower bound are relative to delay constraint and maximum transmission power constraint, respectively. It is observed that the optimal batch size $b^{(t)\star}$ is proportional to the number of edge devices $K$ and the summarization of misalignment error and channel noise error. Also, it deceases with the computation load for executing each data sample, i.e., $C_k$.  Although higher computation speeds $\{\zeta_k\}$ leads to the lower latency, it may reduce the batch size $b^{(t)\star}$ as it induce higher energy cost for each sample. These observations are quite aligned to those in the previous work on ISCC enabled horizontal federated learning \cite{WenTWC25}. Moreover, the sensing power $p_{k,\rm s}^{(t)\star}$ increases with the sensing noise power $\delta_s^2$ for the propose of noise suppression. 
}
\end{remark}

\subsection{Optimization of Denoising Factor and Transmission Power}
Next, we optimize the denoising factor and transmission power under any given sensing power and size of data batch. With the obtained $\left\{p_{k,s}^{(t)},b^{(t)}\right\}$, the objective function in problem (P1) is simplified as 
\begin{align*}
&\tilde{\Omega}\left( \left\{p_{k}^{(t)},\eta^{(t)}\right\}\right) =\notag\\
&\!\sum_{t=1}^{T}\!\frac{\tilde{c}_1}{b^{(t)}}\!\!\left[ \sum\limits_{k\in\mathcal K}\left(\frac{h_{k}^{(t)}\sqrt{p_{k}^{(t)}}}{\sqrt{\eta^{(t)}}} \!\!-1\! \!\right)^2\!\!\!\!+\!\sum\limits_{k\in\mathcal K}\frac{\left(h_{k}^{(t)}\right)^2p_{k}^{(t)}\tilde\delta_{k}^{(t)}}{\eta^{(t)}}+\frac{\sigma_z^2}{\eta^{(t)}}\right]\!\!,\!\!
\end{align*}	
with $ \tilde\delta_{k}^{(t)}=\left( \delta_{k,s}^2+\frac{\delta_s^2}{p_{k,s}^{(t)}}\right)G_2^2$.
and thus the original problem (P1) is reduced as 
\begin{align}
\textrm{(P1.2):}\min\limits_{\left\{p_{k}^{(t)}\geq 0,\eta^{(t)}\geq 0\right\}} &	 \tilde{\Omega}\left( \left\{p_{k}^{(t)},\eta^{(t)}\right\}\right)\notag\\
\textrm{s.t.} ~~~~~&\!\!p_{k}^{(t)}  \leq d b^{(t)}P^{\rm max}_k,\forall k\in{\mathcal K}, t\in\mathcal{T}\label{P12_Power}\\
&\!\!\sum\limits_{t\in\mathcal{T}}\left(p_{k}^{(t)}\tau_{\rm slot} \right)  \leq \tilde{E_k} ,~\forall k\in{\mathcal K},\label{P12_Energy}
\end{align}
where  $\tilde{E_k}= E_k-\sum\limits_{t\in\mathcal{T}}\left(p_{k,s}^{(t)} b^{(t)} \tau_{k, \rm s}^{(t)}+ \kappa_k C_k b^{(t)}\zeta_k^2 \right)$, and constraints \eqref{P12_Power} and \eqref{P12_Energy} are reduced from \eqref{sys_bar_P_max1} and \eqref{Sys_Energy}, respectively.
However, problem (P1.2) is still non-convex under any given sensing power and size of data batch, due to the coupling of the transmission power and denoising factors in the objective function.

To deal with this difficulty, we adopt the alternating optimization technique to problem (P1.2), where the denoising factor $\{\eta^{(t)}\}$ and the transmission power scaling factor $\{p_{k}^{(t)}\} $ are optimized iteratively in an alternating manner, by considering the other to be given in each iteration.

\subsubsection{ Optimization of Denoising Factor}

First, we optimize $\{\eta^{(t)}\}$ in problem (P1.2) under given transmission power scaling factor $\{p_{k}^{(t)}\} $. Therefore, problem (P1.2) can be decomposed into the following $T$ subproblems each for one round $t\in\mathcal{T}$ by dropping the constant $\frac{\tilde{c}_1}{b^{(t)}}$:
\begin{align}
\!\!\!\min_{\eta^{(t)}\geq 0} &\sum\limits_{k\in\mathcal K}\!\!\left(\!\!\frac{h_{k}^{(t)}\sqrt{p_{k}^{(t)}}}{\sqrt{\eta^{(t)}}} \!-1 \!\!\right)^2\!\!\!\!+\!\!\sum\limits_{k\in\mathcal K}\frac{\left(h_{k}^{(t)}\right)^2p_{k}^{(t)}\tilde\delta_{k}^{(t)}}{\eta^{(t)}}+\frac{\sigma_z^2}{\eta^{(t)}}.\!\label{P12_eta}
\end{align}
Then we recast problem \eqref{P12_eta} by introducing an auxiliary variable $\hat{\eta}^{(t)}=1/\sqrt{\eta^{(t)}}$, based on which it is reformulated as 
\begin{align}
\min_{\hat{\eta}^{(t)}\geq 0}~~~ &\sum\limits_{k\in\mathcal K}\!\left(\!h_{k}^{(t)}\sqrt{p_{k}^{(t)}}\hat{\eta}^{(t)} \!-\!1 \right)^2\!+\!\sum\limits_{k\in\mathcal K}\!\!\left(h_{k}^{(t)}\!\right)^2\!p_{k}^{(t)}\tilde\delta_{k}^{(t)}\left(\hat{\eta}^{(t)}\right)^2\notag\\
&+\!\sigma_z^2\left(\!\hat{\eta}^{(t)}\right)^2.\!\label{P12_hat_eta}
\end{align}
Due to the convexity of the objective function in  problem \eqref{P12_hat_eta}, we could obtain the optimal solution defined as $\hat{\eta}^{(t)*}$ by checking its first derivative, based on which we thus accordingly get the optimal solution to problem \eqref{P12_hat_eta} as ${\eta}^{(t)\star}=(\frac{1}{\hat{\eta}^{(t)*}})^2,\forall t\in\cal T$, given in the following proposition.
\begin{proposition}\label{Lemma_Eta}\emph{
	With any given $\left\{p_{k}^{(t)}\right\} $, the optimal solution of ${\eta}_t $ to problem \eqref{P12_eta} is given by
\begin{align}\label{P12_eta_opt}
	{\eta}_t^{\star} =\left( \frac{  \sum\limits_{k\in\mathcal K}\left(h_{k}^{(t)}\right)^2p_{k}^{(t)}\left(\tilde\delta_{k}^{(t)}   +1\right)+\sigma_z^2 }{\sum\limits_{k\in\mathcal K}\left(h_{k}^{(t)}\right)^2p_{k}^{(t)} } \right)^2,~t\in\mathcal{T}.
\end{align}
}
\end{proposition}

%


\subsubsection{Optimization of Transmission Power}

Next, we optimize the transmission power variables $\{p_{k}^{(t)}\}$ in problem (P1.2) under any given denoising factors $\{{\eta}^{(t)}\}$. Therefore, problem (P1.2) is reduced into
\begin{align}
\!\!\!\!\!\! \min\limits_{\left\{p_{k}^{(t)}\geq 0\right\}} &	\!\sum_{t=1}^{T} \frac{\tilde{c}_1}{b^{(t)}}\!\! \left[\!  \sum\limits_{k\in\mathcal K}\!\! \left(\frac{h_{k}^{(t)}\sqrt{p_{k}^{(t)}}}{\sqrt{\eta^{(t)}}}\!\! -\!\! 1 \!\! \right)^2\!\!\! \!\! +\!\! \sum\limits_{k\in\mathcal K}\!\!\frac{\left(\!h_{k}^{(t)}\!\right)^2\!\! p_{k}^{(t)}\tilde\delta_{k}^{(t)}}{\eta^{(t)}}\! \right]\label{SubP_Power}\\
\textrm{s.t.} ~~~&\eqref{P12_Power}~\textrm{and}~\eqref{P12_Energy}.\notag
\end{align}
However, the resultant problem  \eqref{SubP_Power} is still non-convex.  
Via introducing auxiliary variables defined as $ \hat{p}_{k}^{(t)}\triangleq\sqrt{p_{k}^{(t)}}, \forall k\in\mathcal{K}, t\in\mathcal{T}$, problem \eqref{SubP_Power} is equivalently transformed into the following convex form: 
\begin{align}
\!\!\!\!\!\min\limits_{\left\{\hat{p}_{k}^{(t)}\geq 0\right\}} &	 \sum_{t=1}^{T}\!\! \frac{\tilde{c}_1}{b^{(t)}} \!\! \left[ \sum\limits_{k\in\mathcal K}\left(\frac{h_{k}^{(t)}\hat{p}_{k}^{(t)}}{\sqrt{\eta^{(t)}}}\!\!  -\!\! 1\!\!  \right)^2\!\! +\!\! \sum\limits_{k\in\mathcal K}\!\! \frac{\left(\!h_{k}^{(t)}\hat{p}_{k}^{(t)}\!\right)^2\tilde\delta_{k}^{(t)}}{\eta^{(t)}}\!\! \right]\label{SubP_Power_S}\\
\textrm{s.t.} ~~&\hat{p}_{k}^{(t)} \leq \sqrt{\tilde{P}^{\rm max}_k},~\forall k\in{\mathcal K}, ~ t\in\mathcal{T}\label{SubP_MAX_Power}\\
&\sum\limits_{t\in\mathcal{T}}\left(\hat{p}_{k}^{(t)}\right) ^2\tau_{\rm slot}  \leq \tilde{E_k} ,~\forall k\in{\mathcal K},\label{SubP_Energy}
\end{align}
where $\tilde{P}^{\rm max}_k= d b^{(t)}P^{\rm max}_k,~\forall k\in{\mathcal K}, ~ t\in\mathcal{T}$ and 
 constraints \eqref{SubP_MAX_Power} and \eqref{SubP_Energy} follow from \eqref{P12_Power} and \eqref{P12_Energy}, respectively. Notice that problem \eqref{SubP_Power_S} is convex and thus can be optimally solved.
By leveraging the Lagrange duality method, we have the following lemma.

\begin{lemma}\label{lemma_SubP_Power_S}\emph{The optimal solution to problem \eqref{SubP_Power_S} is given as
\begin{align}
&\hat{p}_{k}^{(t)\star}\!=\!\min \!\!\left[\frac{\tilde{c}_1h_{k}^{(t)} \sqrt{\eta^{(t)}}}{ \tilde{c}_1\left(h_{k}^{(t)}\right)^2\!\left(\tilde\delta_{k}^{(t)}   \!+\!1\!\right)\!+\!\alpha_k^{\star}b^{(t)} \eta^{(t)}\tau_{\rm slot} },\sqrt{ \tilde{P}^{\rm max}_k}\!\right]\!.
\label{SubP_hatPower_OPT}
\end{align}
where $\alpha_k^{\star}$ is the optimal dual variable associated with the $k$-th constraint in \eqref{SubP_Energy}.
}
\end{lemma}
\begin{IEEEproof}
This proof is similar to that of Proposition \eqref{SubP_sensing2_Prop}, and thus omitted here due to page limitation.
\end{IEEEproof}
From Lemma~\ref{lemma_SubP_Power_S}, we thus obtain the optimal solution $p_{k}^{(t)\star},
 \forall k\in\mathcal{K},~t\in\mathcal{T}$ to problem \eqref{SubP_Power} as
\begin{align}\label{SubP_Power_Opt}
&p_{k}^{(t)\star}=\left(\hat{p}_{k}^{(t)\star}\right)^2\notag\\
&\!=\!\!\min\!\! \left[\frac{\tilde{c}_1h_{k}^{(t)} \sqrt{\eta^{(t)}}}{ \tilde{c}_1\!\left(h_{k}^{(t)}\right)^2\!\left(\tilde\delta_{k}^{(t)}\!+1\!\right)\!+\!\alpha_k^{\star}b^{(t)} \eta^{(t)}\tau_{\rm slot} },\tilde{P}^{\rm max}_k\right].
\end{align}

\begin{remark}\label{Remark_q}\emph{ It is observed from \eqref{SubP_hatPower_OPT} and \eqref{SubP_Power_Opt} that the optimal transmission power $\left\{p_{k}^{(t)\star}\right\}$ exhibit a regularized channel inversion structure with the regularized term $\alpha_k^{\star}b^{(t)} \eta^{(t)}\tau_{\rm slot} $, which is related to its energy budget in \eqref{SubP_Energy} through the optimal dual variable $\alpha_k^{\star}$. Besides, the transmission power is inversely proportional to the sensing and cluster noises (i.e., $\tilde{\delta}_k^{(t)}$).
In particular, based on the complementary slackness condition for problem \eqref{SubP_Energy}, it follows that 
if $\alpha_k^{\star}>0$ holds for edge device $k\in\mathcal{K}$, then we have $ \sum\limits_{t\in\mathcal{T}}\left(p_{k}^{(t)}\tau_{\rm slot} \right) -\tilde{E_k}=0 $, such that this edge device should run out of its energy; otherwise, if $\alpha_k^{\star}=0$, edge device $k$ should transmit with channel-inversion transmission power control with a coefficient $\tilde{\delta}_k^{(t)}+1$.
}
\end{remark}

Now, with the obtained $\left\{{\eta}^{(t)\star}\right\}$ in \eqref{P12_eta_opt} and $ \left\{p_{k}^{(t)\star}\right\}$ in \eqref{SubP_Power_Opt}, we summarize the complete algorithm to solve problem (P1), in which $\{p_{k}^{(t)}\}$ and $\{{\eta}^{(t)}\}$ are updated alternately in an iterative manner, as shown in Algorithm 1. In each iteration, we first solve problem \eqref{P12_eta} under given $\left\{ p_{k}^{(t)}\right\}$ to update $\left\{{\eta}^{(t)}\right\}$ as $\left\{{\eta}_t^{\star}\right\}$, and then solve \eqref{SubP_Power} under  $\{ \eta_{t}\}$ to update $\{ p_{k}^{(t)}\}$ as $ \{{p}_{k,t}^{\star}\}$. Notice that Algorithm 1 would converge to monotonically non-increasing objective values for problem (P1.2) over rounds. 
Since the optimal value of problem (P1) is monotonically non-increasing at each round. This together with the fact that the optimal value of problem (P1.2) is lower-bounded shows that Algorithm 1 will converge to at least a locally optimal solution to problem (P1.2).

\begin{table}[htp]
\begin{center}
\hrule
\normalsize
\vspace{0.2cm} \textbf{Algorithm 1 for Solving Problem (P1.2)}\vspace{0.2cm}
\hrule \vspace{0.1cm} 
\begin{itemize}
    \item[1]  Input $\{b^{(t)}\}$ and $\{p_{k,\rm s}^{(t),i}\}$.
    \item[2]    Initialization: Set the initial power control $\{p_{k}^{(t),0}\}$ and $i=0$.
    \item[3]  {\bf Repeat:}
                \begin{itemize}
                \item[a)]  With given $ p_{k}^{(t)}=p_{k}^{(t),i}, \forall k\in\mathcal{K}, t\in\mathcal{T}$, obtain the optimal solution to problem \eqref{P12_eta} as $\eta^{(t),i}={\eta}^{(t)\star},~\forall t\in\mathcal{T}$ in \eqref{P12_eta_opt};
                 \item[b)]  With given $\eta^{(t),i}, \forall  t\in\mathcal{T}$, obtain the optimal solution to  problem \eqref{SubP_Power} as $p_{k}^{(t),i}=p_{k}^{(t)\star}, \forall k\in\mathcal{K}, t\in\mathcal{T}$ in \eqref{SubP_Power_Opt};
                \item[c)] Set $p_{k,t}^{(i+1)}=p_{k}^{(t)\star}, \forall k\in\mathcal{K}, t\in\mathcal{T}$, and $i=i+1$.
                \end{itemize}
     \item[4] {\bf Until} the objective value of problem (P1.2) converges within a given threshold.
    \end{itemize}
\hrule \vspace{10pt}
\end{center}
\end{table}

\subsection{Overall Algorithm}

With the obtained solutions of the two subproblems (P1.1) and (P1.2), we adopt an alternating optimization to solve problem (P1), as summarized in Algorithm 2. Problems (P1.1) and (P1.2) are sequentially and iteratively solved via fixing the variables of each other. Notice that the optimal solutions of problem (P1.1) can be achieved and the convergence of Algorithm 1 is guaranteed. This indicates that each step in the iteration leads to a non-increasing objective value and the optimal value of problem (P1) is lower-bounded, Algorithm 2 would converge to a local optimum point.


\begin{table}[htp]
\begin{center}
\hrule
\normalsize
\vspace{0.2cm} \textbf{Algorithm 2 for Solving Problem (P1)}\vspace{0.2cm}
\hrule \vspace{0.1cm} 
\begin{itemize}
    \item[1]  Initialization: Set the initial denoising factor $\{\eta^{(t),0}\}$ and power control $\{p_{k}^{(t),0}\}$ and $i=0$.
    \item[2]  {\bf Repeat:}
                \begin{itemize}
                \item[a)]  With given $\eta^{(t)}=\eta^{(t),j}$ and $ p_{k}^{(t)}=p_{k}^{(t),j}, \forall k\in\mathcal{K}, t\in\mathcal{T}$, obtain the optimal solution to problem (P1.1) as $b^{(t),j}=b^{(t),\star},~\forall t\in\mathcal{T}$ in \eqref{Prop1_b} and $p_{k,\rm s}^{(t),j}=p_{k,\rm s}^{(t)\star}$ in \eqref{SubP_sensing2_Pks};
                 \item[c)]  With given $b^{(t),j}$ and $p_{k,\rm s}^{(t),j}, \forall k\in\mathcal{K},  t\in\mathcal{T}$, obtain the solution to problem (P1.2) as $\eta^{(t)\star}$ and $p_{k}^{(t)\star }, \forall k\in\mathcal{K}, t\in\mathcal{T}$ via Algorithm 1;
                \item[d)] Set $j=j+1$.
                \end{itemize}
     \item[3] {\bf Until} the objective value of problem (P1) converges within a given threshold.
    \end{itemize}
\hrule \vspace{10pt}
\end{center}
\end{table}

\section{Simulation}

This section provides simulation results to validate the learning performance of the proposed design. In the simulation, the wireless channels from edge devices to edge server follow independent and identically distributed (i.i.d.) Rayleigh fading over different rounds, and the path loss is $10^{-3}$. The wireless sensing dataset in \cite{LiuCL22,WenTWC25} is adopted to train ResNet-10, with 4,900,677 model parameters in total. 
Unless otherwise specified, we adopt the following default parameters:  for each device $k\in\mathcal{K}$, the energy threshold is set to be $E_k=1000~\mathrm{J}$, total delay budget as $\Delta_k=300\,\mathrm{s}$, maximum transmit power as $P_k^{\max}=0.05\,\mathrm{W}$, maximum sensing power as $P_{k,s}^{\max}=0.05\,\mathrm{W}$, noise variance as $\sigma_z^2=10^{-9}$, sensing noise variance as $\delta_s^2=10^{-9}$, clutter variance as $\delta_{k,s}^2=10^{-9}$, embedding dimension as $d=100$, per-sample CPU cycles as $C_k\approx 10^{7}$, CPU frequency as $\zeta_k=2\times 10^{9}\,\mathrm{Hz}$, and $T=200$. 
To validate the effectiveness of our proposed scheme, we conduct a seven-class human motion recognition task based on a public wireless sensing dataset \cite{LiuCL22,WenTWC25}, including standing, adult pacing, child pacing, adult walking, child walking, adult walking, and child walking.  And the number of edge devices is $K=3$. The learning rate is set to be 0.1. 

To verify the performance of the proposed ISCC-enabled VFEEL scheme, the following benchmark schemes are considered for performance comparison. 
\begin{itemize}
    \item \textbf{Fixed transmission power:} The transmission power is fixed $p_{k} = 0.5 P_{k}^{\rm max}, \forall k$, while the remaining variables are optimized as in Section \ref{Sec_Opt}.
    \item \textbf{Fixed batch size:} We fix the batch size as $=400, \forall k$, and then optimize the remaining variables as in Section \ref{Sec_Opt}.
    \item \textbf{Fixed denoising factor:} We fix the denoising factor as $\eta = 0.5$, and then optimize the remaining variables as in Section \ref{Sec_Opt}.
\end{itemize}  
 We also compare the proposed scheme with the previous work in ISCC-enabled FEEL work in \cite{WenTWC25}, where the channel-inversion based power control approach is applied to suppress magnitude error induced in AirComp.

\begin{figure*}[htbp]
\centering
\subfigure[Test accuracy]{
\includegraphics[width=0.48\textwidth]{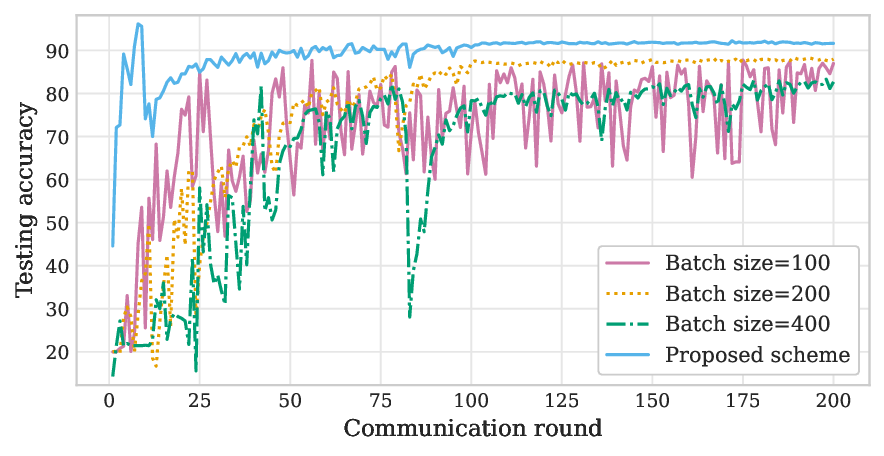}
\label{fig:bt_acc}
}
\subfigure[Training loss]{
\includegraphics[width=0.48\textwidth]{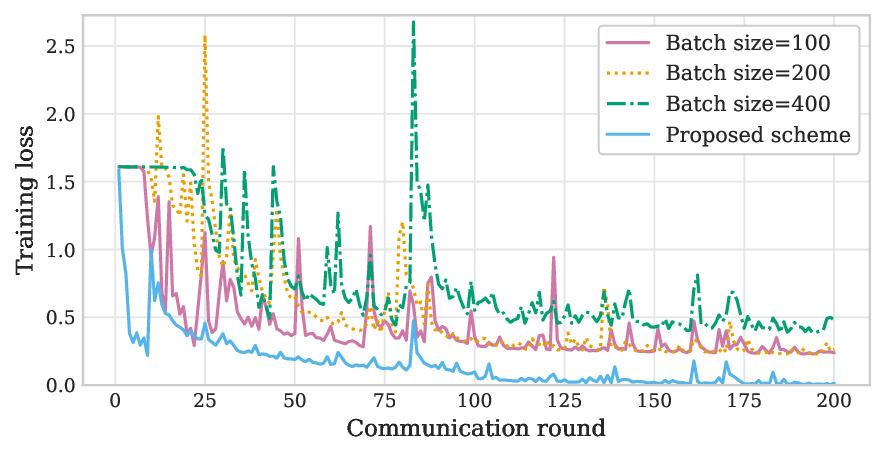}
\label{fig:bt_loss}
}
\caption{Learning performance of ISCC-enabled VFEEL over different batch size, where the batch size is around 150 after optimization in the proposed scheme.}
\label{fig:batch_size_comparison}
\vspace{10pt}
\end{figure*}

Fig. \ref{fig:batch_size_comparison} shows the learning performance (i.e., the test accuracy in Fig. \eqref{fig:bt_acc} and the training loss in Fig. \eqref{fig:bt_loss}) under different batch size with $E_k=3000~\mathrm{J}$ and $\Delta_k=300\,\mathrm{s}, \forall k\in\mathcal{K}$, where the batch size is around 150 after optimization in the proposed scheme.  
It is observed the proposed scheme shows better performance in convergence and accuracy. In particular, when the batch size is lower than the the proposed scheme, the learning performance both in testing accuracy and convergence would be degraded. While increasing the batch size compared with the proposed scheme, although converged, the accuracy would be also affected due to the limited network resource. These observations align with the insight discussed in Remark \ref{Remark_leanding} on the effectiveness of optimization of batch size.


\begin{figure*}[htbp]
\centering
\subfigure[Test accuracy]{
\includegraphics[width=0.48\textwidth]{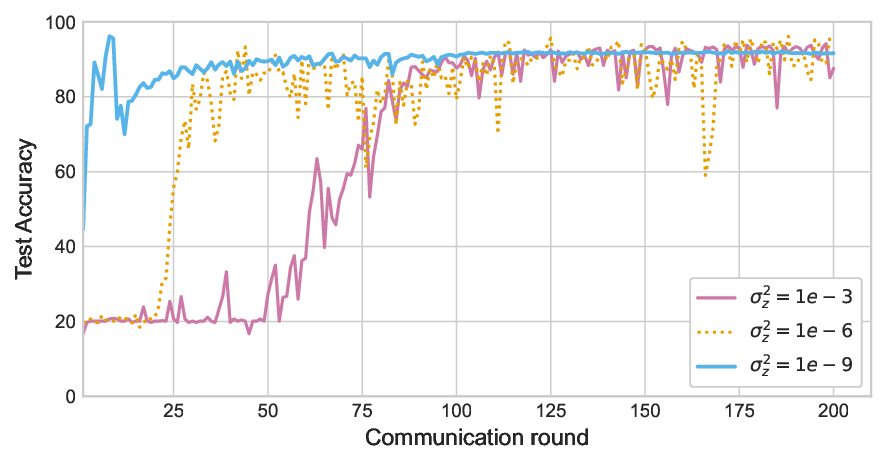}
\label{fig:aircomp_acc}
}
\subfigure[Training loss]{
\includegraphics[width=0.48\textwidth]{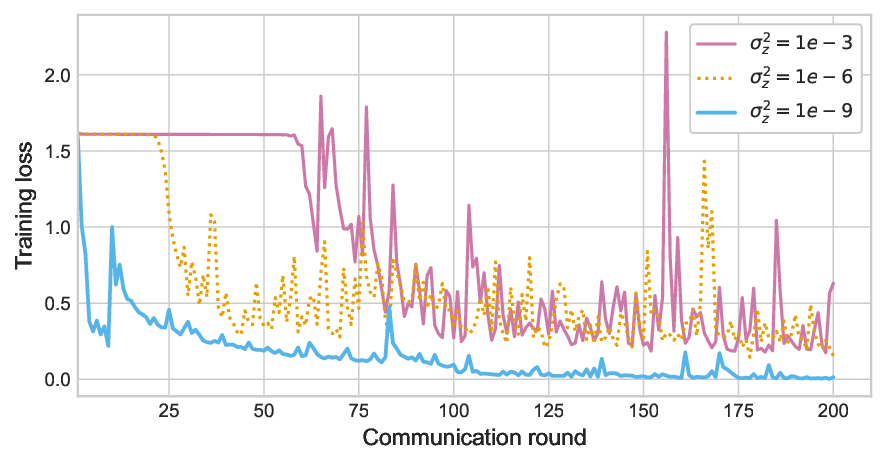}
\label{fig:aircomp_loss}
}
\caption{Learning performance of ISCC-enabled VFEEL over different channel noise variance.}
\label{fig:aircomp_noise_comparison}
\vspace{10pt}
\end{figure*}


Fig. \ref{fig:aircomp_noise_comparison} shows the learning performance (i.e., the test accuracy in Fig. \eqref{fig:aircomp_acc} and the training loss in Fig. \eqref{fig:aircomp_loss}) under different channel noise variances $\sigma_z^2$.  It is observed that a higher testing accuracy and a faster convergence rate are achieved when the channel noise variance is low. Particularly, with larger noise induced, the training loss becomes larger and convergence slows down with fluctuations in accuracy. This shows the effectiveness on the denoising factor optimization to suppress channel noise.




\begin{figure}[t]
  \centering
  \includegraphics[width=0.5\textwidth]{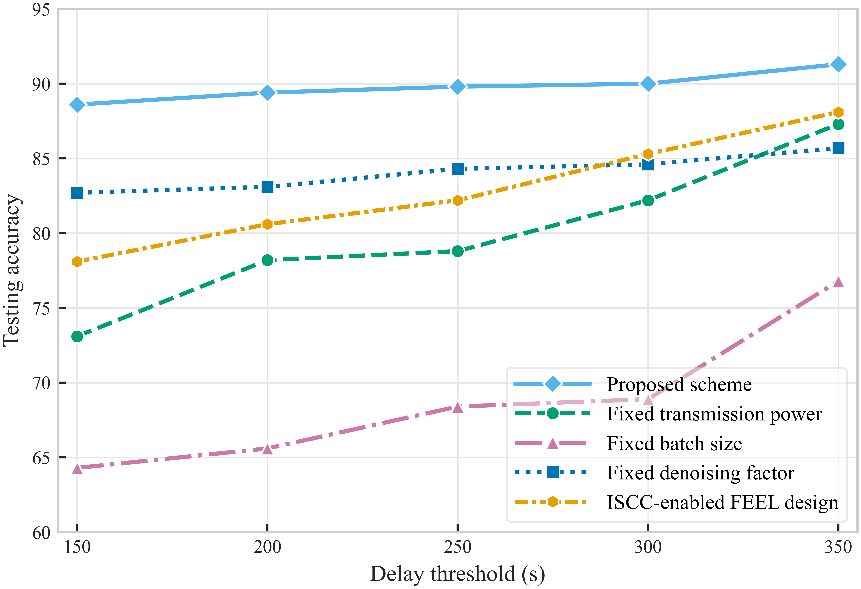}
  \caption{Testing accuracy versus uniform delay threshold $\Delta=\Delta_{k}^{(t)}, \forall k\in\mathcal{K}, t\in\mathcal{T}$.}
  \label{fig:vs_delay}
  \vspace{7pt}
\end{figure}

\begin{figure}[t]
  \centering
  \includegraphics[width=0.45\textwidth]{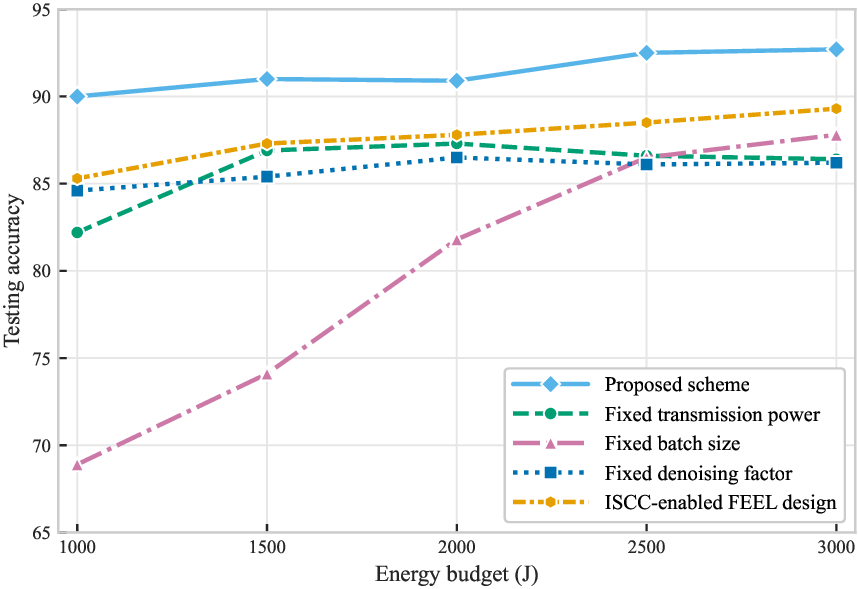}
  \caption{Testing accuracy versus uniform energy budget.}\label{fig:vs_energy}
  \vspace{7pt}
\end{figure}

\begin{figure}[t]
  \centering
  \includegraphics[width=0.45\textwidth]{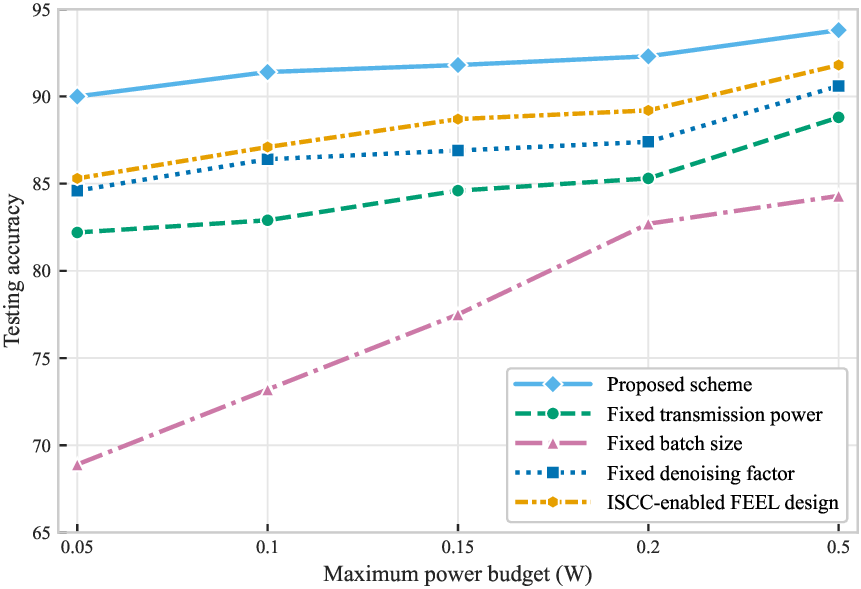}
  \caption{Testing accuracy versus uniform maximum power budget.}\label{fig:vs_pkmax}
  \vspace{7pt}
\end{figure}



Fig. ~\ref{fig:vs_delay} shows the learning performance (namely the test accuracy) versus the delay threshold, where the delay threshold at each edge device is assumed to be uniform, i.e., $\Delta=\Delta_{k}^{(t)}, \forall k\in\mathcal{K}, t\in\mathcal{T}$.
It is observed that under a per-round latency threshold, the proposed scheme outperforms benchmark methods and achieves higher test accuracy when the requirement is satisfied. This demonstrates the advantage of power control optimization in accelerating convergence by mitigating magnitude misalignment error induced by AirComp as well as sensing and noise errors.
Notably, the proposed scheme surpasses the ISCC-based FEEL scheme in \cite{WenTWC25}, which highlights the superiority of the VFEEL framework in leveraging multi-view sensing data.
Performance further improves as the allowable training delay increases, since more samples can be sensed under looser constraints. However, when the delay becomes sufficiently large, performance saturates because energy consumption emerges as the primary bottleneck. 

Fig. ~\ref{fig:vs_energy} shows the test accuracy versus the energy budget, where the energy budget at each edge device is assumed to be uniform, i.e., $\tilde{E}=E_{k}, \forall k\in\mathcal{K}$. A similar trend of initial improvement followed by convergence is observed, as the total energy budget eventually becomes sufficient and no longer serves as the dominant performance bottleneck.

Fig. ~\ref{fig:vs_pkmax} shows the test accuracy versus the maximum power budget, where the maximum power budget at each edge device is assumed to be uniform, i.e.,  $\tilde{P}^{\max}=P_{k}^{\max}, \forall k\in\mathcal{K}$. 
It is observed that the proposed scheme can achieve better learning performance compared with other benchmarking schemes. And increasing the maximum transmission power limit yields a more significant improvement in test accuracy. This also shows the importance of power control optimization.


\section{Conclusion}
This paper considered an ISCC-enabled VFEEL system, where edge devices collected sensing data via wireless signals and fed them into the local model. The resulting embeddings are then transmitted to the edge server via AirComp for efficient aggregation and global model training.
We first analyzed the convergence behavior of the ISCC-enabled VFEEL in terms of the loss function degradation in the presence of sensing noise and aggregation distortions during AirComp. Then, to accelerate the convergence, the batch size, sensing power, and transmission power control at edge devices as well as the denoising factors at edge server were jointly optimized under limited network constraints. To deal with the tight coupling of variables, we proposed an alternating algorithm to efficiently obtain a high-quality solution.
Numerical results validated the learning performance gain achieved by the proposed ISCC-enabled VFEEL scheme compared with other benchmarking schemes. 
How to extend this framework into decentralized scenarios by leveraging the consensus mechanism is quite interesting for future work.

\section*{Acknowledgment}
The authors would like to thank the developers of Qwen, an advanced large language model, for providing valuable assistance during the manuscript preparation process.

\appendix

\subsection{Proof of Lemma~\ref{Lemma_Embed}}\label{Theo_Lemma_Embed}

Let $ \hat{\bm g}_i\left({\bm \theta}_{k}^{(t)}\right) $, $\forall k\in\{0\}\cup\mathcal{K}$ define the gradient of $i$-th sample by combining \eqref{Sys_ChainRule_Server} and \eqref{Sys_ChainRule}, and it is recast by   
\begin{align*}
	 \hat{\bm g}_i\left({\bm \theta}_{k}^{(t)}\right) &= \nabla_{{\bm \theta}_{k}^{(t)}} {\psi}_{k} ( {\bm \theta}_{k}^{(t)};{\bm \xi}_{k,i}^{(t)} ) \nabla_{ {\psi}_{k}} f_i\left({\bm\theta}_0^{(t)}; {\bm \varepsilon}_i^{(t)}+{ \bm \psi}_i^{(t)} \right)\label{Ana-gradient-ki},
\end{align*}
where it follows 
\begin{align*}
&\nabla_{ {\psi}_{k} } f_i\left({\bm\theta}_0^{(t)}; {\bm \varepsilon}_i^{(t)}+{ \bm \psi}_i^{(t)} \right)\triangleq \notag\\
\!&\left\{\begin{array}{ll}
	    \nabla_{{  \psi}_{0} }  f_i\left({\bm\theta}_0^{(t)}; {\bm \varepsilon}_i^{(t)}+{ \bm \psi}_i^{(t)} \right), k=0	\\
		\nabla_{ {\psi}_{k} ( {\bm \theta}_{k}^{(t)};{\bm \xi}_{k,i}^{(t)} ) } \tilde{ \bm \psi}_{i}^{(t)} \nabla_{\tilde{ \bm \psi}_{i}^{(t)} }  f_i\left({\bm\theta}_0^{(t)}; {\bm \varepsilon}_i^{(t)}+{ \bm \psi}_i^{(t)} \right), \forall k\in\mathcal{K}
	\end{array}
	\right.
\end{align*}
Then, we apply Tayler series expansion to $\nabla_{ {\psi}_{k} } f_i \left(  {\bm \varepsilon}_i^{(t)}+{ \bm \psi}_i^{(t)} \right) $, $\forall k\in\{0\}\cup\mathcal{K}$ at the point ${\bm \psi}_i^{(t)} $, given by 
\begin{align*}
&\nabla_{ {\psi}_{k} } f_i\left({\bm\theta}_0^{(t)}; {\bm \varepsilon}_i^{(t)}+{ \bm \psi}_i^{(t)} \right) \notag\\
&= \nabla_{ {\psi}_{k}} f_i\left({\bm\theta}_0^{(t)}; { \bm \psi}_i^{(t)} \right)+\left(\nabla_{ {\psi}_{k}}^2 f_i\left({\bm\theta}_0^{(t)};{\bm \psi}_{i}^{(t)} \right)\right)^{\dagger} {\bm \varepsilon}_i^{(t)}+O({\psi}_{k} ),
\end{align*}
where $O({\psi}_{k} )$ is the infinitesimal of higher order and is ignored in the sequential analysis. Thus, the gradient of $i$-th sample is reformulated as 
\begin{align*}
	&\hat{\bm g}_i\left({\bm \theta}_{k}^{(t)}\right) = \nabla_{{\bm \theta}_{k}^{(t)}} {\psi}_{k} ( {\bm \theta}_{k}^{(t)};{\bm \xi}_{k,i}^{(t)} ) \nabla_{ {\psi}_{k}} f_i\left({\bm\theta}_0^{(t)}; { \bm \psi}_i^{(t)} \right)\notag\\
   &~~~~~~~~~~~~~+ \nabla_{{\bm \theta}_{k}^{(t)}} {\psi}_{k} ( {\bm \theta}_{k}^{(t)};{\bm \xi}_{k,i}^{(t)} )\left(\nabla_{ {\psi}_{k}}^2 f_i\left({\bm\theta}_0^{(t)};{\bm \psi}_{i}^{(t)} \right)\right)^{\dagger} {\bm \varepsilon}_i^{(t)}\notag\\
	&\!=\!\! \nabla_{k}\!f_i\!\left(\!{\bm\theta}_0^{(t)};{\bm \psi}_{i}^{(t)}\! \right)\!\!+\!\!\nabla_{{\bm \theta}_{k}^{(t)}} {\psi}_{k} ( {\bm \theta}_{k}^{(t)};{\bm \xi}_{k,i}^{(t)} )\!\left(\!\nabla_{ {\psi}_{k}}^2 f_i\left({\bm\theta}_0^{(t)};{\bm \psi}_{i}^{(t)} \!\right)\!\!\right)^{\dagger}\! {\bm \varepsilon}_i^{(t)}\!.
\end{align*}
According to Assumption \ref{Assum_VarianceBound} and $ \hat{\bm g}\left({\bm \theta}_{k}^{(t)}\right)=\frac{1}{b^{(t)}}\sum_{i=1}^{b^{(t)}} \hat{\bm g}_i\left({\bm \theta}_{k}^{(t)}\right),\forall k\in\{0\}\cup\mathcal{K}$, we have 
\begin{align}
	\mathbb{E}\left[ 	\hat{\bm g}\left({\bm \theta}_{k}^{(t)}\right)\right]&=\nabla_k F({\bm \Theta}^{(t)} ),\forall k\in\{0\}\cup\mathcal{K}, t\in\mathcal{T},
\end{align}
which holds due to \eqref{Ana_AggBias}.
Based on this result, the variance of  $	\hat{\bm g}\left({\bm \theta}_{k}^{(t)}\right)$ is given by
\begin{align}
&\mathbb{E} \left\|	\hat{\bm g}\left({\bm \theta}_{k}^{(t)}\right)-\nabla_k F({\bm \Theta}^{(t)} )\right\|^2\notag\\
&=\mathbb{E} \left\|	\frac{1}{b^{(t)}}\sum_{i=1}^{b^{(t)}} \hat{\bm g}_i\left({\bm \theta}_{k}^{(t)}\right)-\nabla_k F({\bm \Theta}^{(t)} )\right\|^2\notag\\
&=\mathbb{E}\! \left\|\!\frac{1}{b^{(t)}}\!\sum_{i=1}^{b^{(t)}} \!\nabla_{{\bm \theta}_{k}^{(t)}} {\psi}_{k} ( {\bm \theta}_{k}^{(t)};{\bm \xi}_{k,i}^{(t)} )\left(\!\nabla_{\!\! {\psi}_{k}}^2 f_i\left(\!{\bm\theta}_0^{(t)};{\bm \psi}_{i}^{(t)} \right)\right)^{\dagger}\! {\bm \varepsilon}_i^{(t)}\!\right\|^2\!\!\notag\\
&+\mathbb{E} \left\|	\frac{1}{b^{(t)}}\sum_{i=1}^{b^{(t)}}  \nabla_{{\bm \theta}_{k}^{(t)}}f_i\left({\bm\theta}_0^{(t)};{\bm \psi}_{i}^{(t)} \!\right)\!-\!\nabla_k F({\bm \Theta}^{(t)} )\right\|^2,\label{Ana-gradient-expectation}
\end{align}
in which \eqref{Ana-gradient-expectation} holds as the expectation of its cross terms equals to zero.
According to Assumption \ref{Assum_VarianceBound}, the first term in \eqref{Ana-gradient-expectation} is bounded by
\begin{align*}
\mathbb{E} \left\|\!	\frac{1}{b^{(t)}}\!\sum_{i=1}^{b^{(t)}}  \!\nabla_{\!{\bm \theta}_{k}^{(t)}}\!f_i\left({\bm\theta}_0^{(t)};{\bm \psi}_{i}^{(t)} \right)\!-\!\nabla_k F({\bm \Theta}^{(t)} )\!\right\|^2\!\leq \frac{\sigma^2}{b^{(t)}}\!.
\end{align*}
Under Assumptions \eqref{Assump_BoundedHessian} and \eqref{Assump_BoundedEmbedGra}, the squared norm of the partial derivatives w.r.t. ${\bm \theta}_k$ of edge device $k$'s embedding multiplied by the Tayler expansion term is bounded by 
\begin{align}
	&\left\|\nabla_{{\bm \theta}_{k}^{(t)}} {\psi}_{k} ( {\bm \theta}_{k}^{(t)};{\bm \xi}_{k,i}^{(t)} )\left(\nabla_{ {\psi}_{k}}^2 f_i\left({\bm\theta}_0^{(t)};{\bm \psi}_{i}^{(t)} \right)\right)^{\dagger} {\bm \varepsilon}_i^{(t)}\right\|^2\notag\\
	&\leq\left\|\nabla_{{\bm \theta}_{k}^{(t)}} {\psi}_{k} ( {\bm \theta}_{k}^{(t)};{\bm \xi}_{k,i}^{(t)} )\right\|_{\mathcal{F}}^2 \left\|\left(\nabla_{ {\psi}_{k}}^2 f_i\left({\bm\theta}_0^{(t)};{\bm \psi}_{i}^{(t)} \right)\right)^{\dagger} {\bm \varepsilon}_i^{(t)}\right\|_{\mathcal{F}}^2\notag\\
     &\leq G_1^2 \left\|\nabla_{ {\psi}_{k}}^2 f_i\left({\bm\theta}_0^{(t)};{\bm \psi}_{i}^{(t)} \right) \right\|_{\mathcal{F}}^2 \left\|{\bm \varepsilon}_i^{(t)}\right\|^2\label{Ana_gradient-Tayler-Err1}\\
      &\leq G_1^2 \Psi^2 \left\|{\bm \varepsilon}_i^{(t)}\right\|^2\label{Ana_gradient-Tayler-Err2},
\end{align}
where \eqref{Ana_gradient-Tayler-Err1} and \eqref{Ana_gradient-Tayler-Err2} follow Assumptions \eqref{Assump_BoundedEmbedGra} and  \eqref{Assump_BoundedHessian}, respectively.
Thus, the second term in \eqref{Ana-gradient-expectation} is bounded by 
\begin{align}
	&\mathbb{E} \left\|\frac{1}{b^{(t)}}\sum_{i=1}^{b^{(t)}} \nabla_{{\bm \theta}_{k}^{(t)}} {\psi}_{k} ( {\bm \theta}_{k}^{(t)};{\bm \xi}_{k,i}^{(t)} )\left(\nabla_{ {\psi}_{k}}^2 f_i\left({\bm\theta}_0^{(t)};{\bm \psi}_{i}^{(t)} \right)\right)^{\dagger} {\bm \varepsilon}_i^{(t)}\right\|^2\notag\\
	&\!\leq \!\frac{1}{\left(\!b^{(t)}\right)^2}\!\sum_{i=1}^{b^{(t)}} \mathbb{E}\ \left\|\!\nabla_{{\bm \theta}_{k}^{(t)}} {\psi}_{k} \!( {\bm \theta}_{k}^{(t)};{\bm \xi}_{k,i}^{(t)} )\!\left(\!\nabla_{ {\psi}_{k}}^2 f_i\!\left(\!{\bm\theta}_0^{(t)};{\bm \psi}_{i}^{(t)}\! \right)\!\!\right)^{\dagger}\! {\bm \varepsilon}_i^{(t)}\!\right\|^2\!\notag\\
	&\leq   \frac{G_1^2 \Psi^2}{\left(b^{(t)}\right)^2}\sum_{i=1}^{b^{(t)}} \mathbb{E}\left[\left\|{\bm \varepsilon}_i^{(t)}\right\|^2\right]\label{Ana_gradient-expectation_Err2}.
\end{align}
Therefore, the variance of $\hat{\bm g}\left({\bm \theta}_{k}^{(t)}\right)$ is recast as 
\begin{align}
	\mathbb{E}\!\!\left\|\hat{\bm g}\!\left({\bm \theta}_{k}^{(t)}\!\right)\!\!-\!\!\nabla_k F({\bm \Theta}^{(t)} )\right\|^2\!&\!\leq \! \frac{\sigma^2}{b^{(t)}}+\frac{G_1^2 \Psi^2}{\left(b^{(t)}\right)^2}\sum_{i=1}^{b^{(t)}} \mathbb{E}\left\|{\bm \varepsilon}_i^{(t)}\right\|^2\notag\\
	&\leq \!\! \frac{\sigma^2+G_1^2 \Psi^2}{b^{(t)}} \mathbb{E}\left\|\tilde{\bm \varepsilon}^{(t)}\right\|^2\!\!,\label{Ana_Exp_Err1}
\end{align}
where \eqref{Ana_Exp_Err1} holds following the MSE of aggregation error in \eqref{Ana_AggMSE3}.
This thus completes the proof.

\subsection{Proof of Lemma~\ref{Lemma_Loss_Reduction}}\label{Lemma_Loss_Reduction_Proof}

The proof follows by relating the norm of the gradient to the expected improvement made at each communication round.
Based on \eqref{Sys-ModelUpdate}, the updating rule of training model could be rewrote as 
\begin{align}
	{\bm \Theta}^{(t+1)}={\bm \Theta}^{(t)}-\mu^{(t)} \hat{\bm g}\left({\bm \Theta}^{(t)}\right),
\end{align}
where $\hat{\bm g}\left({\bm \Theta}^{(t)}\right)=\left[\hat{\bm g}\left({\bm \theta}_{0}^{(t)}\right)^{\dagger},\cdots,\hat{\bm g}\left({\bm \theta}_{K}^{(t)}\right)^{\dagger}\right]^{\dagger}$.

Based on Assumption \ref{Assump_Smooth}, it follows that
\begin{align}
	&F\left({\bm \Theta}^{(t+1)}\right)- F\left({\bm \Theta}^{(t)}\right)\notag\\
    &\leq 
	\left(\nabla F\left({\bm \Theta}^{(t)}\right) \right)^{\dagger} \left({\bm \Theta}^{(t+1)}-{\bm \Theta}^{(t)}\right) + \frac{L}{2}\left\|{\bm \Theta}^{(t+1)}-{\bm \Theta}^{(t)}\right\|^2\notag\\
	&\!=\!\! -\! \left(\!\nabla F\left({\bm \Theta}^{(t)}\right) \!\!\right)^{\dagger}\! \left(\!\mu^{(t)} \hat{\bm g}\left({\bm \Theta}^{(t)}\!\right)\!\right) \!+ \!\frac{L}{2}\left\|\mu^{(t)} \hat{\bm g}\left({\bm \Theta}^{(t)}\right)\right\|^2\!\label{App1_GradientUp}\\
	&\!= \!- \!\mu^{(t)} \left(\nabla F\left({\bm \Theta}^{(t)}\right)\! \right)^{\dagger}\! \hat{\bm g}\!\left(\!{\bm \Theta}^{(t)}\!\right) \!+\! \frac{L(\mu^{(t)})^2}{2}\!\left\| \hat{\bm g}\!\left(\!{\bm \Theta}^{(t)}\!\right)\!\right\|^2\label{Proof_Smoth1}\!,\!\!
	\end{align}
where \eqref{App1_GradientUp} follows the updating rule of gradients. By taking expectation at both sides of \eqref{Proof_Smoth1}, we have
\begin{align}
&\mathbb{E}\left[F\left({\bm \Theta}^{(t+1)}\right)- F\left({\bm \Theta}^{(t)}\right) \right]\notag\\
&\leq\! -\! \mu^{(t)} \!\left(\!\nabla F\left({\bm \Theta}^{(t)}\!\right)\! \right)^{\dagger} \!\mathbb{E}\!\left[ \hat{\bm g}\left({\bm \Theta}^{(t)}\right)\right] \!+ \!\frac{L(\mu^{(t)})^2}{2}\!\mathbb{E}\!\left[\!\left\| \hat{\bm g}\left(\!{\bm \Theta}^{(t)}\!\right)\!\right\|^2\!\right]\!\notag\\
&=\frac{L(\mu^{(t)})^2}{2}\mathbb{E}\left[\left\| \hat{\bm g}\left({\bm \Theta}^{(t)}\right)-\nabla F\left({\bm \Theta}^{(t)}\right)+\nabla F\left({\bm \Theta}^{(t)}\right)\right\|^2\right]\notag\\
&~~~~- \mu^{(t)} \left\|\nabla F\left({\bm \Theta}^{(t)}\right) \right\|^{2} \notag\\
&= - \mu^{(t)} \left\|\nabla F\left({\bm \Theta}^{(t)}\right) \right\|^{2}  +\frac{L(\mu^{(t)})^2}{2}\mathbb{E}\left[\left\| \nabla F\left({\bm \Theta}^{(t)}\right)\right\|^2\right]\notag\\
&~~~+\frac{L(\mu^{(t)})^2}{2}\mathbb{E}\left[\left\| \hat{\bm g}\left({\bm \Theta}^{(t)}\right)-\nabla F\left({\bm \Theta}^{(t)}\right)\right\|^2\right]\label{Proof1_E1}\\
&= \left(- \mu^{(t)}+\frac{L(\mu^{(t)})^2}{2}\right) \left\|\nabla F\left({\bm \Theta}^{(t)}\right) \right\|^{2} \notag\\
&~~~+ \frac{L(\mu^{(t)})^2}{2}\sum_{k=0}^{K}\mathbb{E}\left[\left\| \hat{\bm g}\left({\bm \theta}_{k}^{(t)}\right)-\nabla_k F\left({\bm \Theta}^{(t)}\right)\right\|^2\right]\notag\\
&\leq \left(- \mu^{(t)}+\frac{L(\mu^{(t)})^2}{2}\right) \left\|\nabla F\left({\bm \Theta}^{(t)}\right) \right\|^{2}  \notag\\
&~~~+ \frac{L(\mu^{(t)})^2}{2}\sum_{k=0}^{K}\left[\frac{\sigma^2+G_1^2 \Psi^2}{b^{(t)}} \mathbb{E}\left[\left\|\tilde{\bm \varepsilon}^{(t)}\right\|^2\right]\right]\label{Proof1_E2},
\end{align}
where both \eqref{Proof1_E1} and \eqref{Proof1_E2} are obtained according to Lemma \ref{Lemma_Embed}.

\subsection{Proof of Theorem~\ref{Theo_Gradient}}\label{Theo_Gradient_Proof}

Applying the assumption of fixed learning  $\mu=\mu^{(t)}, \forall t\in\mathcal{T}$ with $0\leq \mu<\frac{2}{L}$, the expected per-round loss descent is given by
\begin{align*}  
&\left\|\nabla F\left({\bm \Theta}^{(t)}\right) \right\|^{2}  \leq \frac{2\mathbb{E}\left[F\left({\bm \Theta}^{(t)}\right)- F\left({\bm \Theta}^{(t+1)}\right) \right]}{\mu\left(2-L\mu\right)} \notag\\
&~~~~~~~~~~~~~~~~~~~~~+ \frac{L\mu(K+1)(\sigma^2+G_1^2 \Psi^2)}{\left(2-L\mu\right)b^{(t)}} \mathbb{E}\left\|\tilde{\bm \varepsilon}^{(t)}\right\|^2.
\end{align*}
Summing over all training rounds $t=0,\cdots, T-1$, we have 
\begin{align*} 
&\sum_{t=1}^{T}\left\|\nabla F\left({\bm \Theta}^{(t)}\right) \right\|^{2}  \leq \frac{2\mathbb{E}\left[F\left({\bm \Theta}^{(1)}\right)- F\left({\bm \Theta}^{(T+1)}\right) \right]}{\mu\left(2-L\mu\right)} \notag\\
&~~~~~~~~~~~~~~~~~~~+ \sum_{t=1}^{T}\frac{L\mu(K+1)(\sigma^2+G_1^2 \Psi^2)}{\left(2-L\mu\right)b^{(t)}} \mathbb{E}\left\|\tilde{\bm \varepsilon}^{(t)}\right\|^2.
\end{align*}
Taking average of the above inequality, we have \eqref{Theo_Gradient_Eq}.
This completes the proof.

\subsection{Proof of Lemma~\ref{SubP_sensing2_Prop}}\label{Proof_SubP_sensing2_Prop}

Recall that problem \eqref{SubP_sensing2} is convex and satisfies the Slater's condition. The strong duality thus holds between problem \eqref{SubP_sensing2} and its Lagrange dual problem \cite{cvx}. 
By leveraging the Lagrange duality method, we can optimally solve problem \eqref{SubP_sensing2}.

Let $\lambda_k\ge 0$ denote the dual variable associated with  the $k$-th constraints in \eqref{SubP_sensing2C1}.
Then the partial Lagrangian of problem \eqref{SubP_sensing2} is
\begin{align*}
&\mathcal{L}_1\!\left(\!\left\{\!e_{k,s}^{(t)}, b^{(t)}\!\right\}\!\right)\!=\!\tilde{c}_1  \sum\limits_{t\in\mathcal{T}}\!\frac{A_1^{(t)}}{b^{(t)}} \!+\! \tilde{c}_1 G_2^2\sum\limits_{t\in\mathcal{T}}\!\sum\limits_{k\in\mathcal K}\!\frac{\left(h_{k}^{(t)}\right)^2p_{k}^{(t)}\delta_s^2}{\eta^{(t)}e_{k, \rm s}^{(t)}} \notag\\
&+\sum \limits_{k\in\mathcal{K}} \lambda_k \left(\sum \limits_{t\in\mathcal{T}}\left(e_{k, \rm s}^{(t)} \tau_{k, \rm s}^{(t)}+ \kappa_k C_k b^{(t)}\zeta_k^2+p_{k}^{(t)}\tau_{\rm slot} \right) -E_k\right).
\end{align*}
Then the dual function is
\begin{align}\label{Dual_Sensing2}
W_1(\{\lambda_k\})=\min_{\left\{e_{k,s}^{(t)}\geq 0,b^{(t)}\geq 0\right\}} ~&\mathcal{L}_1\left(\left\{e_{k,s}^{(t)}, b^{(t)}\right\}\right)\\
{\rm s.t.}~~~~~~&\!b^{(t)}
\leq \min_{k\in\mathcal{K}} \tilde\Delta_{k}^{(t)},\forall t\in\mathcal{T}\label{Dual_sensing2C2}\\ 
&\!\!b^{(t)} \!\!\geq\! \max_{k\in\mathcal{K}}\!\frac{p_{k}^{(t)} }{d P^{\rm max}_k},\forall t\in\mathcal{T}\!\! \label{Dual_sensing2C3},
\end{align}
where $\tilde\Delta_{k}^{(t)}\triangleq \frac{\Delta_{k}^{(t)}}{ \left(\tau_{k, \rm s}^{(t)}+\frac{C_k }{\zeta_k}+\frac{d \tau_{\rm slot}}{M} \right)}, ~\forall k\in\mathcal{K},~\forall t\in\mathcal{T}$, constraints \eqref{Dual_sensing2C2} and \eqref{Dual_sensing2C3} are reduced from \eqref{SubP_sensing2C2} and \eqref{SubP_sensing2C3}, respectively.

Accordingly, the dual problem of problem \eqref{SubP_sensing2} is given as
\begin{align}
\mathbf{D1:} \min_{\{\lambda_k\ge 0\}} ~&W_1(\{\lambda_k\}).
\end{align}
Due to the fact that the strong duality holds between problems \eqref{SubP_sensing2} and (D1), we can solve problem \eqref{SubP_sensing2} by equivalently solving its dual problem (D1). For notational convenience, let $\left\{e_{k,s}^{(t)\star}, b^{(t)\star}\right\}$ denote the optimal primal solution to problem \eqref{SubP_sensing2}, and $\{\lambda_k^{\star}\}$  denote the optimal dual solution to problem (D1).

Next, we first evaluate the dual function $W_1(\{\lambda_k\})$ under any given feasible $\{\lambda_k\}$, and then obtain the optimal dual variables $\{\lambda_k^{\star}\}$ to maximize $W_1(\{\lambda_k\})$.
First, we obtain $W_1(\{\lambda_k\})$ by solving problem \eqref{Dual_Sensing2} under any given feasible $\{\lambda_k\}$. 
By checking the first-order derivation of the objective function in problem \eqref{Dual_Sensing2}, we have the following lemma. 

\begin{lemma}\label{lemma_Dual_Sensing2}\emph{The optimal solution to problem \eqref{Dual_Sensing2} denoted by $\{e_{k,s}^{(t)*},b^{(t)*}\}$ is given as
\begin{align}
e_{k,s}^{(t)*}&=\sqrt{ \frac{ \tilde{c}_1G_2^2  \left(h_{k}^{(t)}\right)^2p_{k}^{(t)}\delta_s^2 }{\lambda_k\tau_{k, \rm s}^{(t)} \eta^{(t)} } }, ~\forall k\in{\mathcal K},~\forall t\in\mathcal{T}\\
b^{(t)*}&= \left(\sqrt{ \frac{ \tilde{c}_1A_1^{(t)} }{ \sum \limits_{k\in\mathcal{K}}\lambda_k \kappa_k C_k \zeta_k^2  }}    \right)_{b^{(t),\rm l}}^{b^{(t),\rm u}},~\forall t\in\mathcal{T},
\end{align}
where $( x)_{u_1}^{u_2}\triangleq\min(u_2,\max(u_1,x))$ with $b^{(t),\rm l}=\max \limits_{k\in\mathcal{K}}\frac{p_{k}^{(t)} }{d P^{\rm max}_k}$ and $b^{(t),\rm u}=\min \limits_{k\in\mathcal{K}} \tilde\Delta_{k}^{(t)}$.
}
\end{lemma}
Therefore, with Lemma \ref{lemma_Dual_Sensing2},  problem \eqref{Dual_Sensing2} is solved, and the dual function $W_1(\{\lambda_k\})$ is accordingly obtained. It next to obtain the optimal  $\{\lambda_k\}$.
Since the dual function $W_1(\{\lambda_k\})$ is concave and non-differentiable, the ellipsoid method \cite{Gradient} is applied to obtain $\{\lambda_k^{\star}\}$. For the objective function in \eqref{Dual_Sensing2}, the subgradient w.r.t. $\lambda_k$ is $ \sum \limits_{t\in\mathcal{T}}\left(e_{k, \rm s}^{(t)*} \tau_{k, \rm s}^{(t)}+ \kappa_k C_k b^{(t)*}\zeta_k^2+p_{k}^{(t)}\tau_{\rm slot} \right) -E_k$.
By replacing $\{\lambda_k\}$ in Lemma~\ref{lemma_Dual_Sensing2} with $\{\lambda_k^{\star}\}$, the optimal solution to problem \eqref{SubP_sensing2} is accordingly obtained as shown in Proposition~\ref{SubP_sensing2_Prop}. This thus completes the proof.

\bibliography{AirCompforFL}
\bibliographystyle{IEEEtran}


\end{document}